\documentclass{kais}

\usepackage{wrapfig,psfig,epsf}
\usepackage[dcucite]{harvard}

\usepackage{wrapfig,psfig,epsf}
\usepackage{enumerate}
\usepackage[labelformat=parens,labelsep=quad,skip=3pt]{subfig}
\usepackage{graphicx}
\usepackage{booktabs}
\usepackage{arydshln}
\usepackage{multirow}
\usepackage{multicol}
\usepackage{rotating}
\usepackage{romannum}
\usepackage{algorithmic}
\usepackage[ruled]{algorithm2e}

\usepackage{amsfonts}
\usepackage{amsmath}
\usepackage{multirow}
\usepackage{cite}

\usepackage{float}
\restylefloat{table}
\restylefloat{figure}

\newtheorem{definition}{Definition}
\newtheorem{problem}{Problem}

\accepted{29 April 2019 - Computers in Biology and Medicine - This is the preface version}

\pubyear{2019}

\begin{document}
\pagenumbering{arabic}
\label{firstpage}

\title{MCP: a Multi-Component learning machine to Predict protein secondary structure}
\author[Leila Khalatbari et al.]{
	Leila Khalatbari{\small $~^{\dag}$}, MR Kangavari{\small $~^{\dag}$}, Saeid Hosseini{\small $~^{\copyright\dag\S}$}, Hongzhi Yin{\small $~^{\#}$}, Ngai-Man Cheung $~^{\S}$\\$~^{\textbf{\dag}}$ School of Computer Engineering, Iran University of Science and Technology, Tehran, Iran\\
	$~^{\S}$ ST Electronics - SUTD Cyber Security Laboratory, Singapore University of Technology and Design, Singapore\\
	$~^{\textbf{\#}}$ School of Information Technology and Electrical Engineering, University of Queensland, Brisbane, Australia\\
	$~^{\textbf{\copyright}}$ Corresponding Author}
 
\maketitle
\vspace{-4mm}
\begin{abstract}
	The Gene or DNA sequence in every cell does not control genetic properties on its own; Rather, this is done through translation of DNA into protein and subsequent formation of a certain 3D structure. The biological function of a	 protein is tightly connected to its specific 3D structure. Prediction of the protein secondary structure is a crucial intermediate step towards elucidating its 3D structure and function. Traditional experimental methods for prediction of protein structure are expensive and time-consuming. Therefore, various machine learning approaches have been proposed to predict the protein secondary structure. Nevertheless, the average accuracy of the suggested solutions has hardly reached beyond 80\%. The possible underlying reasons are the ambiguous sequence-structure relation, noise in input protein data, class imbalance, and the high dimensionality of the encoding schemes that represent the protein sequence. In this paper, we propose an accurate multi-component prediction machine to overcome the challenges of protein structure prediction. We devise a multi-component designation to address the high complexity challenge in sequence-structure relation. Furthermore, we utilize a compound string dissimilarity measure to directly interpret protein sequence content and avoid information loss. In order to improve the accuracy, we employ two different classifiers including support vector machine and fuzzy nearest neighbor and collectively aggregate the classification outcomes to infer the final protein secondary structures. We conduct comprehensive experiments to compare our model with the current state-of-the-art approaches. The experimental results demonstrate that given a set of input sequences, our multi-component framework can accurately predict the protein structure. Nevertheless, the effectiveness of our unified model can be further enhanced through framework configuration. 
\end{abstract}


\vspace{-5mm}
\section{Introduction}
\label{sec:intro}

\begin{figure}[H]
	\centering
	\includegraphics[scale=0.7]{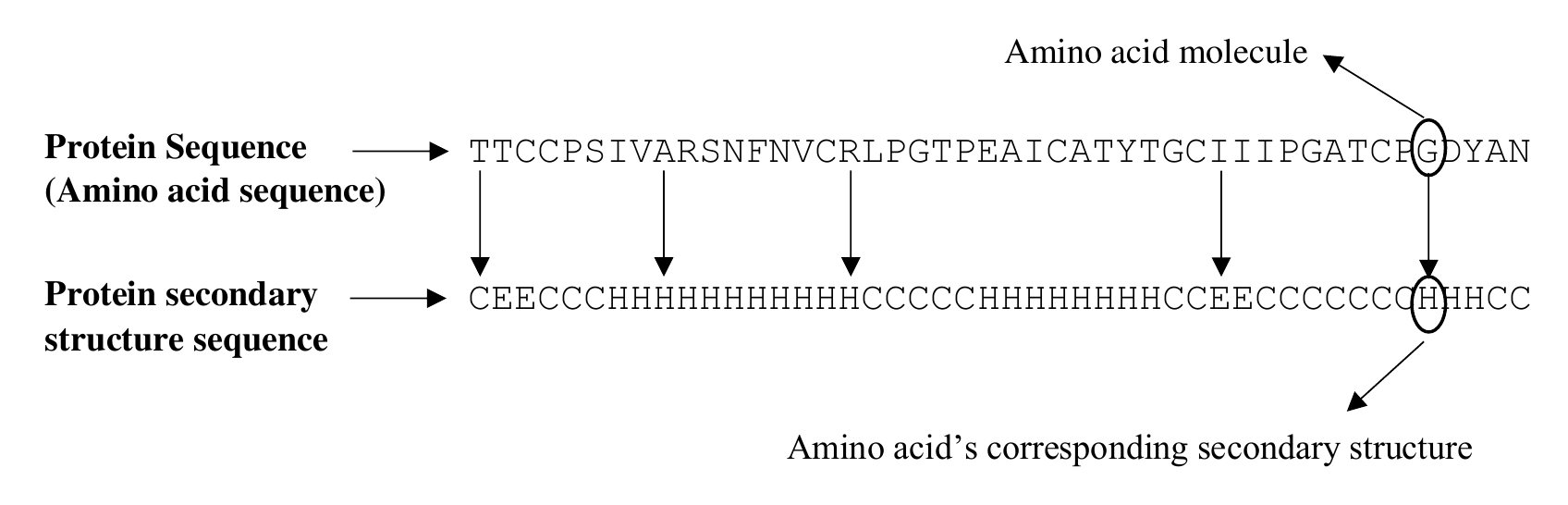}
	\caption{sequence-structure mapping}
	\vspace{-6mm}
	\label{fig:seq-struct}
\end{figure}

In this paper, we focus on interpreting of the sequential string data which is crucial to many applications including bioinformatics and molecular biology. Accordingly, this interpretation problem is studied in the context of computational biology. Prediction of protein secondary structure from the input string sequences is the ground to foster protein function determination and design better drugs \cite{Wang2016, Tan2015, Fang2018}. Hence, understanding of the protein sequences to predict protein structure has become an important use-case in molecular biology. Given protein sequences composed of amino-acid molecules, our aim is to predict the secondary structure that each amino-acid adopts through a classification paradigm. In Figure \ref{fig:seq-struct}, the top string chain illustrates a protein sequence. Each letter of this sequence represents an amino-acid molecule. Our aim is to assign each amino-acid molecule to one of three classes of protein secondary structure named as $\alpha$-helix (H), $\beta$-sheet (E) and \textit{coil} (C).
\vspace{-6mm}
\vspace{-4mm}
Template-based methods and machine learning strategies are two computational approaches for prediction of protein secondary structure. Template-based methods neither yield higher accuracy compared to machine learning methods nor perform well on non-homologous proteins \cite{Lin2010}. As a consequence, effective machine learning-based strategies are much more preferable. Feature extraction from protein sequences is the first step when applying a machine learning approach. However, the extracted features may not reflect all the information a sequence contains and thus can lead to information loss \cite{Li2017,Liu2010,Chen2017}. Nevertheless, from biological perspective, protein sequence contains indispensable information to adopt certain structures \cite{Lin2010,Liu2010,Liu2017}. While predicting such structures is an appealing task, challenges abound. First, the relationship between a sequence and its corresponding structure is quite complex \cite{Wang2016,Krissinel2007}. Second, the selected features dramatically influence the learner's effectiveness \cite{Liu2010}. Additionally, the training data including protein sequences and their related known structures are partially noisy. Lastly, known as class imbalance phenomena, the amino-acid samples are not distributed equally in three classes of the protein structures \cite{Alirezaee2012}.\\
To address the challenges, we devise a Multi-Component Predictor (MCP) which is capable to directly process amino-acid sequences into accurate protein structures. While every component participates in the enhancement and correction of the prediction results, the multi-component property of our solution can learn various information from the input protein sequences. The MCP framework processes the textual context of protein sequences which yields three advantages. First, it avoids information loss through processing the primary sequence data and bypassing the feature extraction procedure. Second, our proposed framework can eliminate the negative effects of certain feature subsets that can further promote the effectiveness of the learner. Third, Given the input protein sequences, the MCP framework can interpret a latent natural language via employing of the dissimilarity measures. Such latent language can reveal hidden relationships between protein sequences.\\
Our proposed framework employs two efficient algorithms of Support Vector Machine (SVM) and Fuzzy K-Nearest Neighbor (FKNN) in parallel. The \textit{edit} distance function forms the \textit{edit} kernel for SVM that infers the dissimilarity among input sequences. Furthermore, we embed a compound dissimilarity measure called $\widetilde{d}$ into FKNN module. $ \widetilde{d} $ works based on n-gram scores, LZ scores, and a new parameter called dissimilarity rate ($ \rho_d $). The output of each learner passes through a filtering component to refine the biologically meaningless structures. The corrected output from filtration enters the aggregation pool that can further make a consensus among the decisions of edit-SVM and $ \widetilde{d}$-FKNN .\\
A worthwhile advantage of the proposed method is its extensive flexibility which promotes the development of more accurate and advanced versions based on the primary solution. For instance, one might want to add more classifiers and expand the ensemble size with different base learners. It is also possible to utilize the dissimilarity measure ($ \widetilde{d}$) in SVM kernel  and subsequently analyze the outcomes more in details. One might pass numerical features (i.e. protein sequence profiles) to a specific learner and the string sequences to another to collectively investigate the resultant differences. Fuzzification of the SVM module can enhance the aggregation process and consequently enrich the final prediction results. Also, a weighted form of the compound dissimilarity measure can lead to accuracy enhancement. Finally, our proposed solution can incorporate a dynamic parameter optimization module which can significantly improve the effectiveness of the prediction results. \\
Our contributions in this study are threefold:
\vspace{-2mm}
\begin{itemize}
	\item We devise a flexible multi-component prediction framework (MCP) which can directly process the latent contexts of the input protein sequences and suggest accurate output secondary structures. Our model can be further generalized to perform on an arbitrary number of classes.
	\item We employ two different classifiers of edit-SVM and $ \widetilde{d}$-FKNN and collectively aggregate the filtered classification outcomes to infer the final prediction results. 
	\item  We achieve better accuracy in prediction of protein secondary structure using various modules of classification, aggregation and dissimilarity measures.
\end{itemize}
The remainder of this paper is organized as follows. In section 2, we briefly summarize related work in the literature of protein secondary structure prediction. In section 3, we provide the preliminary definitions, define problems and  present the overview of our framework. In section 4, we elucidate the procedure of building the proposed model and provide the underlying techniques. We also prove that are model can be generalized to perform on an arbitrary number of classes. In section 5, we examine the effectiveness of the competitor baselines and report the experimental results. In section 6, we offer a conclusion and discuss the future work.
\clearpage
\vspace{-4mm}
\section{Related Work}
\label{sec:relWork}
\vspace{-3mm}
Predicting the structure using a set of string sequences has been well-studied in biology \cite{Garnier1978, Rost1993, Bondugula2005, Lin2010, Johal2014, Fang2018, Fang2018 }. Neglecting external knowledge-bases \cite{hosseini2014location}, machine learning-based methods such as ensemble models, and deep learning approaches have been recently exploited to increase the accuracy of the prediction results. In this section, we discuss the related work in two major aspects: First, \textit{Statistical and mining methods}, Second, \textit{Machine Learning-based models}. Table \ref{table:literature}, briefly demonstrates an overview of the literature. 
\vspace{-3mm}
\begin{table}[h]
	\footnotesize
	\centering
	\caption{An overview of the literature of protein secondary structure prediction}
	\label{table:literature}
	\begin{tabular}{rp{15.07em}l}
		\toprule
		\multicolumn{1}{p{10.5em}}{\textbf{Category of methods}} & \textbf{methods} & \multicolumn{1}{p{10em}}{\textbf{Reference}} \\
		\midrule
		\multicolumn{1}{l}{\multirow{5}[2]{*}{Probabilistic and mining }} & Chou-Fesman & \cite{Chou1974} \\
		& GOR   & \cite{Garnier1996}  \\
		& Hidden Markov Model (HMM) & \cite{Bystroff2008, Chen2007}\ \\
		& Decision-tree &  \cite{Mossos2014, He2006}  \\
		& Natural Language Processing (NLP) & \cite{Lin2010}  \\
		& Distance-based learners & \cite{Ghosh2008, Bondugula2005, Tan2015} \\
		\midrule
		\multicolumn{1}{l}{\multirow{3}[2]{*}{Machine learning }} & Neural networks and deep learning & \cite{Babaei2012, Babaei2010, Alirezaee2012, Dinubhai2014, Wang2008, Paliwal2015, Spencer2015, Wang2016} \\
		& Support vector machines (SVM) & \cite{Zangooei2012, Zangooei2012a, Nguyen2003, Lin2010a}\\
		& Multi-component approaches &  \cite{Chen2007, Dinubhai2014, Wang2008, Yaseen2014, Patel2014, Alirezaee2012, Babaei2012, Johal2014, Babaei2010}\\
		& & \cite{Nguyen2003, Wang2016, Zangooei2012, Zangooei2012a, He2006, Spencer2015, Bouziane2015, Zamani2015}\\
		\bottomrule
	\end{tabular}%
	\label{tab:addlabel}%
\end{table}%

\textbf{Probabilistic and mining methods:}
Primary probabilistic methods \cite{Chou1974,Garnier1996} are based on empirical analytics  and mainly compute the tendency of each amino-acid in protein sequence to form a particular secondary structure (i.e. probabilities are calculated based on the frequency of each amino-acid in each secondary class). GOR \cite{Garnier1996} extends Chou-Fesman \cite{Chou1974} to improve prediction performance through including the context of each amino-acid. In probabilistic terms, it computes the conditional probability for each amino-acid to adopt a certain secondary structure, given that its neighbors have formed that structure. Since the structure of an amino-acid is correlated with its neighbors, in this paper, we also use a sliding window to incorporate the neighboring context in the prediction process. A more recent probabilistic approach for protein secondary structure prediction is Hidden Markov Model (HMM)\cite{Bystroff2008, Chen2007}. The HMM graphical models well adapt to one-dimensional sequence processinsg. When applying HMMs, the states of the graph will be secondary classes and structures are determined using output probabilities of HMM \cite{Bystroff2008}. Chen et al. \cite{Chen2007} employ Markov Model of the third order (as a feature extraction method) to generate a sequence encoding scheme that is subsequently fed into the SVM to predict the protein structure. In our approach, secondary structures are decided from probabilities that are calculated through the fuzzy membership functions.
\\
Tree-based methods are also used in mining approaches \cite{Wu2008}. Mossos et al. \cite{Mossos2014} extracts the rules from FS-Tree - with a modified support - to leverage sequence-structure mappings. In another tree-approach, \cite{He2006} applies SVM to eliminate the noise and the outlier data. He et al. \cite{He2006} firstly pass the refined data to a decision tree. Subsequently, they predict the proteins structures using the extracted rules from the decision tree. Moreover, the NLP-based methods \cite{Hua2012} consider the textual context of the input sequences to infer the missing structures. \cite{Lin2010} exploit n-gram patterns to create a dictionary of synonymous words that can be later used to compute sequence similarities. In our work, we also employ a compound measure including an n-gram metric which estimates the dissimilarities among sequence chunks. Nevertheless, the distance-based classifiers have also been employed in structure prediction \cite{Ghosh2008}. K-Nearest Neighbor algorithm (KNN) and its fuzzified versions \cite{Bondugula2005, Tan2015} are the most popular distance-based learners. Similarly our multi-component framework takes advantage of the fuzzified KNN.\\
\textbf{Machine learning-based approaches:}
In the field of sequence-structure mapping, there are three major groups of machine learning models: \textit{neural networks and deep learning paradigms}, \textit{support vector machines} and \textit{multi-component learners}.
The Neural Networks (NN) are the first generation of machine-learning approaches that are used for protein structure prediction. In practice, employing a well-decided architecture of neural networks leads to a fine estimation of class boundaries. The latest and the most effective version of neural networks is Deep Neural Networks (DNN). Deep networks learn different levels of information abstraction through multiple hidden layers \cite{Paliwal2015}. The mainstream deep networks comprise recurrent neural networks \cite{Babaei2012, Babaei2010}, feedforward multilayer perceptron \cite{Dinubhai2014} and deep convolution neural fields \cite{Paliwal2015}. According to the literature \cite{Masulli2009, Babaei2012, Pollastri2007}, different versions of recurrent neural networks greatly suit processing sequence data. For an instance, bidirectional recurrent neural networks are capable to utilize the information along the entire sequence. While time is a multi-aspect entity \cite{Hosseini2017,Hosseini2017a,Hosseini,hosseini2018exploiting}, the long short-term recurrent neural networks can also retain the information over long periods of time \cite{Paliwal2015}. Deep convolutional neural fields involves more sequence information in its learning process and takes into account the interdependencies of the adjacent context\cite{Wang2016}. The more recent studies both in shallow \cite{Alirezaee2012, Wang2008} and deep neural networks \cite{Spencer2015}, combine the predictions of a number of such networks in an ensemble fashion. Spencer et al. \cite{Spencer2015} propose an ensemble of three DNNs with a cascade architecture. The model is trained using the restricted Boltzmann machine that works with the real-valued data and the contrastive divergence. Despite the strength of neural networks, the choice of proper architecture parameters such as the number of neurons, layers, and activation functions remains an issue. This can significantly affect the prediction outcome. Moreover, there is a chance that the algorithm falls into a local minima. Since, the support vector machines can resolve the parameter selection and the local minima issues, we employ the SVM component in our framework. \\
SVMs are among the most accurate learners in the literature of protein secondary structure prediction \cite{Bouziane2015, Zangooei2012a}. Because of the optimization nature, SVM models perform  more accurately than NNs in many applications \cite{Bouziane2015, Zangooei2012a}. However SVM kernel should be tuned properly. Zangooei et al. \cite{Zangooei2012, Zangooei2012a} use a dynamic weight allocation function to assign weights to three ubiquitous kernels and later fuse them into a single kernel. Furthermore, a parallel hierarchical grid search is applied to tune the kernel parameters. According to \cite{Zangooei2012}, converting the classification to a regression problem and then employing Support Vector Regression (SVR) can further enhance the prediction accuracy. Therefore, aiming to determine the final protein structure, \cite{Zangooei2012} utilizes a non-dominated sorting genetic algorithm to map the real-valued SVR outputs to integer values. The SVM models are also employed as ensemble components \cite{Nguyen2003, Lin2010a}. Nguyen et al. \cite{Nguyen2003} develop an SVM-based cascade architecture where the second layer produces the final prediction results through combining the outputs from the first layer. \\
Multi-component methods employ a variety of complementary modules to unveil the relationship between the input and output vectors. Some categories of such computational modules are various learners, optimization strategies, distance or dissimilarity measures, evolutionary algorithms, and etc. In practice, the multi-component machines perform more competently than single learners. The reason relies on the fact that each component can overcome a part of the challenge. Several methods reviewed in this section were developed in an ensemble or multi-component manner. The first group of multi-component approaches \cite{Chen2007, Dinubhai2014, Wang2008, Yaseen2014, Patel2014} employ complementary modules beside the learning algorithms to promote the prediction results. Similarly, aiming to foster prediction accuracy, the second category \cite{Alirezaee2012, Babaei2012, Johal2014, Babaei2010, Nguyen2003} exploit multiple classifiers of the same type with various features. The third class of Multi-component approaches \cite{Wang2016, Zangooei2012, Zangooei2012a, He2006, Spencer2015, Bouziane2015, Zamani2015} combine classifiers of different types \cite{abachi2018statistical} \cite{maskouni2018auto} that are at times equipped with the complementary components. As every module in a multi-component framework is able to enhance or rectify a learner's prediction outcomes, in this work we devise a multi-component framework to better address the challenges of the secondary structure prediction.
\vspace{-6mm}
\section{Problem Statement}
\label{sec:buildModel}
\vspace{-4mm}
In this section, we offer primary concepts, notations, and the framework overview.
\vspace{-10mm}
\subsection{Preliminary concepts}
\label{subsec:priliCon}
\vspace{-4mm}
Since we study the problem of knowledge extraction from sequential string data in the context of molecular biology, we commence with biological concepts. The Gene or DNA sequence in every cell does not control genetic properties on its own; rather, this is done through the translation of DNA sequence into protein and formation of a certain structure. Hence, the proteins are the functional units of the cells whose functions are tightly connected to their structure. 
\vspace{-2mm}
\begin{definition}
	(protein sequence) A protein sequence $P_{i}$, of the length $l$, is a string composing of $l$ amino-acids (i.e. $P_{i}\left[ m \right]$) participating in protein's formation. Each amino-acid molecule in a protein sequence is represented by an alphabetic letter ($P_{i}\left[ m \right]\in \sum$). There are 20 different types of amino-acids in nature ($\vert\sum\vert = 20$).
\end{definition}
\vspace{-4mm}
\begin{definition}
	(protein secondary structure) The secondary structure of a protein is formed by the local compositions of neighboring amino-acids through peptide bonds. During this chemical reaction, the element of water is removed and what is left of the amino-acid molecules is called amino-acid residues. Thus we refer to amino-acid residue as residue from hereafter. Every residue $P_{i}\left[ r \right]$ is assigned with a secondary structure (i.e. $S_{i}\left[ r \right]$). There are three classes of protein secondary structure, named as $\alpha$-helix, $\beta$-sheet, and the \textit{coil}. The secondary structures are represented by three letters. Therefore, $S_{i}\left[ r \right]\in\lbrace H, E, C\vert H\equiv\alpha-helix, E\equiv\beta-sheets$, and $C\equiv coils\rbrace$. Hence, a protein sequence with $l$ characters will correspond to a sequence of secondary structure with the length of $l$. Each character in the structure sequence (i.e. $S_{i}\left[ r \right]$) is associated with its corresponding amino-acid (i.e. $P_{i}\left[ m \right]$) in the primary protein sequence.
\end{definition}
\vspace{-4mm}
\begin{definition}
	(the dissimilarity rate) the dissimilarity rate $\rho_{d}$ is the number of unique non-identical characters divided by the number of unique identical characters that a pair of sequences share. We disregard the position of characters.
\end{definition}
\vspace{-2mm}
The secondary structure of a residue is strongly influenced by the type of its neighboring residues \cite{Lee2009}. Therefore, to effectively predict a residue's secondary structure, its neighbors must be involved. Hence, we employ the simple but effective sliding window approach to include the adjacent residues in the prediction process. Accordingly, the residue whose structure is going to be predicted will be placed in the middle of the window.

\vspace{-7mm}
\subsection{Problem definition}
\label{subsec:problemDef}
\vspace{-3mm}
Given a set of protein sequences $P_{i}$, our aim is to model the mapping $f:P_{i} \rightarrow S_{i}$. Therefore, we infer the similarity between the sequences in $P_{i}$ via a compound dissimilarity measure. Concurrently, to address the complexity of the function $f$, we devise our unified framework as a multi-component learning machine. 
\vspace{-2mm}
\begin{problem}
	(sequence similarity inference) Given a set of protein sequences $P_{i}$, our aim is to infer similarity between each pair of sequences in $P_{i}$ through a Compound Dissimilarity measure (CD). 
\end{problem}
\vspace{-3mm}
\begin{problem}
	(multi-component learning) Given our compound dissimilarity measure (CD) and the protein sequences $P_{i}$, our goal is to devise a multi-component learning machine to take $P_{i}$ as input and consume protein sequence dissimilarities. Each module of the multi-component machine is expected to enhance the prediction accuracy of the secondary structure (i.e. $S_{i}$) and facilitate an effective aggregation among the decisions of the classifiers. 
\end{problem}
\vspace{-8mm}
\subsection{Framework Overview}
\label{subsec:FrameOver}
\begin{figure}[h]
	\begin{frame}{}
		\centering
		\includegraphics[trim=0cm 0cm 0cm 0cm,clip=true,scale=0.42]{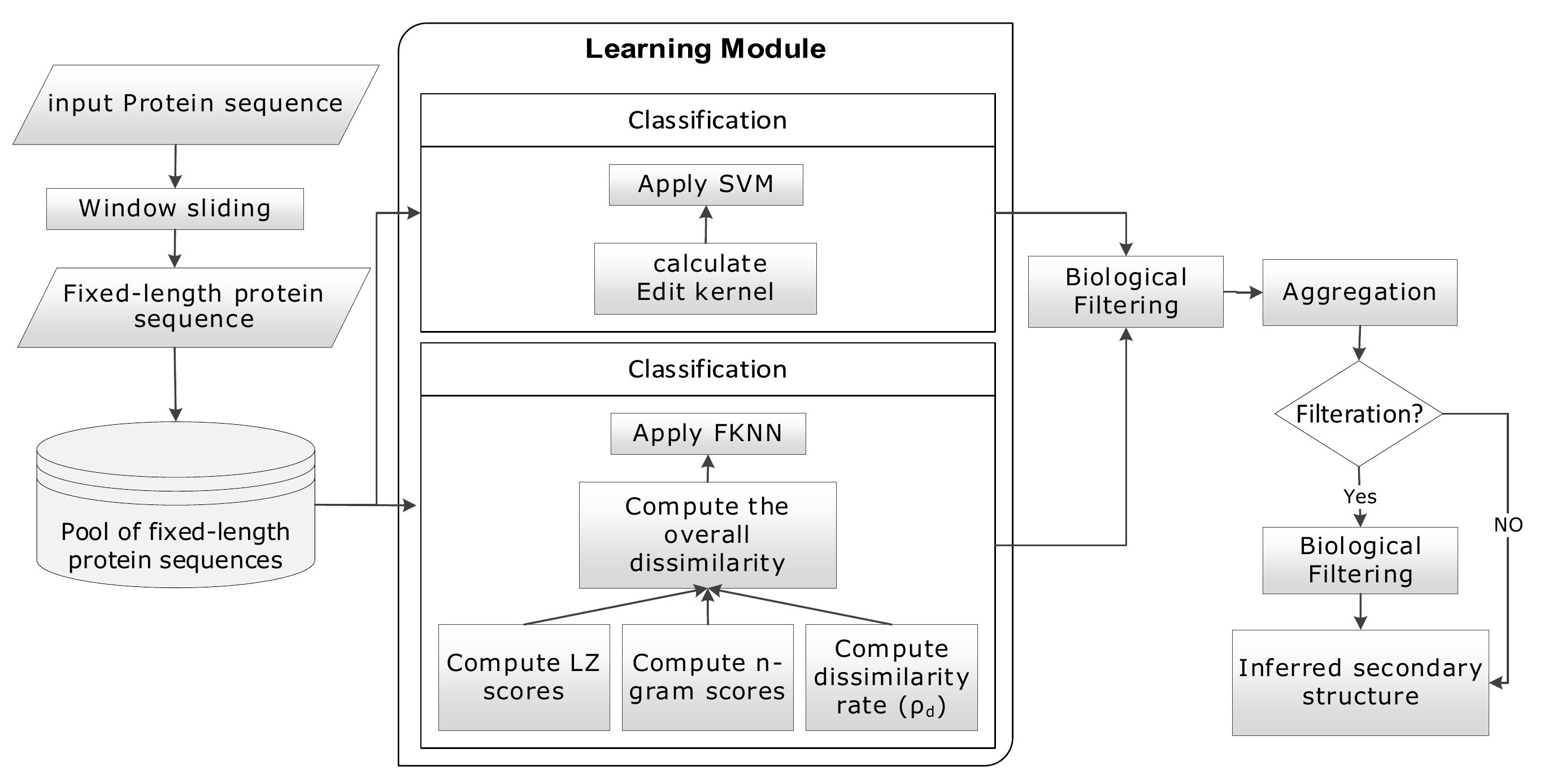}
		\vspace{-4mm}
		\caption{The framework of our proposed approach}
		\label{fig:frameOver}
	\end{frame}
\end{figure}

\vspace{-4mm}
Figure \ref{fig:frameOver} illustrates our multi-component framework for prediction of protein secondary structure from the input protein sequences.
Since each component contributes to error correction and enhances the prediction accuracy, the multi-component framework can better address the mapping procedure ($f$). The strength of a multi-component learning machine mainly stems from the diversity of its components and the effectiveness of its aggregation method. Consequently, we utilize a pair of structurally diverse classifiers (SVM and FKNN) to form the learning core.\\
\vspace{-10mm}
\vspace{-4mm}
\\Initially, a \textit {sliding window} of size $h$ chunks the input protein sequences into the fixed-length set of strings ($P^h$). Including $h-1$ neighbors of the central residue from the sliding window involves the long-range interactions among amino-acids. These interactions are a valuable information for prediction of protein secondary structure.\\
In the \textit{learning phase}, to infer the sequence-structure mapping (i.e. $f$), the set of $P^h$ is fed into two parallel classifiers, $\widetilde{ d }$-FKNN and edit-SVM. As these classifiers are capable to directly learn from $P^h$, we bypass the feature extraction process.
The $\widetilde{ d }$-FKNN processes $P^h$ using the compound dissimilarity measure named as $\widetilde{ d }$ which is computed using three parameters of LZ-scores, n-gram scores and the dissimilarity rate (i.e. $\rho_{d}$) between sequence pairs ($p_{i}^h,p_{j}^h\in P^h$). Each of the three composing elements of $\widetilde{ d }$ infers the sequence dissimilarities from a different aspect. LZ complexity scores are able to explore sequence order information, as well as repeated patterns or the degree of randomness \cite{Tan2015,Liu2010}. From another perspective, the n-gram scores can capture local similarities between sequences which reflect sequence variations during evolution \cite{Lin2010}. Furthermore, the dissimilarity rate reflects the extent of the difference in the type of amino-acid molecules composing each $p_{i}^h$. Fusing these scores into a dissimilarity measure can better infer the sequence-structure relation. The \textit{edit} kernel enables SVM to effectively handle string sequence data (Section \ref{sec:evaluation}). Since $\widetilde{ d }$-FKNN and edit-SVM learn in parallel, better efficiency is provided. The output of each learning module passes through a \textit{filtering} component to eliminate the biologically meaningless structures. In the \textit{aggregation pool}, five various aggregation rules are accommodated. Each aggregation rule makes a consensus between the decisions of $\widetilde{ d }$-FKNN and edit-SVM in a different fashion. The fuzzy property of KNN can further enhance the aggregation process. The final secondary structure (i.e. $S_{i}$) can be obtained through filtering of the aggregation results.\\
Q. What necessity for sequence processing? the effective measure of dissimilarity in sequence processing has a few advantages over processing of the extracted numerical feature vectors. First, it prevents information loss from the rich protein sequences. Second, the learner's performance varies on different sets of numerical features and it is not always feasible to find the optimal feature set. Additionally, numerical features or encoding schemes (i.e. numerical representations for string protein sequences) may lead to high dimensional feature vectors which can negatively affect the learner's performance \cite{Liu2010}. From the biological perspective, every protein sequence contains all essential information for structure adoption \cite{Lin2010,Liu2010,Liu2017}. Hence, we utilize the compound dissimilarity measure ${\widetilde{ d }}$ alongside with the \textit{edit} kernel to process the string sequences effectively.
\vspace{-7mm}
\section{Related material}
\label{sec:RM}
\vspace{-4mm}
In this section we intend to introduce the whole procedure of secondary structure prediction. After that we briefly describe the utilized data sets. Then the evaluation metrics and the test train methods are introduced. Ultimately a source for access to codes and data sets are provided. 
\vspace{-7mm}
\subsection{Step-by-step prediction procedure}
\label{sec:predProc}
\vspace{-4mm}
To construct an effective sequence-based secondary structure predictor there are five steps to go over carefully as followed in []:\\
\Romannum{1}) To select valid and appropriate test and train benchmark data sets which are recent and used by many other methods in the related field. The data sets must satisfy some conditions. For instance, the proteins should not be homologous, there must be enough number of proteins to draw valid conclusions from the results and more than one data set should be employed to verify the outcome. A popular data set provides a comprehensive comparative framework with the proposed method.\\
\Romannum{2}) To properly encode the input biological sequence to further feed to a predictor. The encoded format must preserve as much information as possible in comparison with the original data and also have high correlation with the target label. It is important to mention that the input representation greatly affects the performance of predictor and consequently the final results. Hence the representation of protein sequences for a predictor is a challenge and must be thoughtfully addressed.\\
\Romannum{3}) To design a powerful prediction engine to effectively address the challenges of secondary structure prediction. The more robust, accurate and efficient the engine is, the more it is effective.\\
\Romannum{4}) To select an appropriate evaluation framework including proper and comprehensive evaluation measures and methods to adapt to the conditions of the problem. In case of secondary structure prediction as a classification problem, accuracy, specificity, sensitivity, MCC and SOV are most frequently used and well describe the effectiveness of a method. Also cross validation is popular for tuning parameters of the problem as well as training and testing the prediction method. \\
\Romannum{5}) To establish a public user-friendly web-server which predicts the secondary structure of the user's input protein data on the basis of the proposed method. 
Through out this paper, we elaborate each step in detail.\\
\vspace{-10mm}
\subsection{Test, train and Data set description}
\label{sec:dataset}
\vspace{-4mm}
RS126, CASP12, CASP11, table of properties.\\
To validate the effectiveness of our approach three publicly available data sets of RS126, CASP11 and CASP12 are employed. It is important for a data set to have been used to evaluate various methods as comparison of methods is merely possible on the identical data sets. RS126 data set is widely used through out the literature of the problem. Also to evaluate our method against more recent approaches and the latest proteins with more complications the two recent CASP11 and CASP12 are utilized.\\
To accomplish a comprehensive validation we first employ RS126 for training and testing via 10-fold cross validation and report the results. Then we utilize CASP11 and CASP12 as two recent Independent test sets to further confirm our achievements.\\
Table \ref{table:data set} provides some quantities of the introduced data sets. \\
\vspace{-1mm}
\begin{table}[H]
	\footnotesize
	\centering
	\caption{Data sets description}
	\vspace{-2mm}
	\label{table:data set}
	\begin{tabular}{p{7em}l:l:l:l}
		\midrule
		\multicolumn{1}{p{5em}}{\textbf{Data set}} & \multicolumn{1}{p{8em}}{\textbf{No of proteins }} & \multicolumn{1}{p{8 em}}{\textbf{Sequence length}} & \multicolumn{1}{p{6.355em}}{\textbf{No of residues}} & \multicolumn{1}{p{4.215em}}{\textbf{Year of creation}} \\
		\midrule
		RS126 \cite{Rost1993} & 126 & 185 & 32465 & '1993'\\[4pt]
		CASP11 & 85 & 44 to 669 & 20498 & 2014 \\[4pt]
		CASP12 & 40 & 75 to 670 & 10526 & 2016 \\[4pt]
		\bottomrule
	\end{tabular}%
	\label{tab:addlabel}%
\end{table}%
\vspace{-5mm}

The RS126 dataset contains globular non-homologous proteins. The \textit{coil} with a portion of 45\% is the most common structure in the dataset, while the other 23\% and 32\% of the residues are respectively categorized as \textit{$\beta$-sheet} and \textit{$\alpha$-helix}. \\
According to the CASP official definition, CASP11 contains proteins which are regarded as hard targets meaning that it is difficult to detect their homologous structure templates from known protein structures. Also, the reported accuracy for CASP12 structures is usually lower than other CASP data sets since it contains proteins that are hard to classify.\\
It is worthwile to mention that as RS126 is far older than CASP11 and CASP12, there no common targets in these train and test data sets. \\
\vspace{-10mm}
\subsection{Evaluation measures}
\label{sec:measures}
\vspace{-4mm}
To demonstrate the real effectiveness of a method it is important to employ various evaluation metrics. The reason lies in the fact that a method might show higher values for some measures and possibly much lower values for some other. A truly effective and reliable method is the one that shows high and concurrently balanced values for various measures. Therefore to comprehensively evaluate our approach, we employed 5 widely used evaluation measures for protein secondary structure predictors as well as general classifiers. These measures include overall accuracy ($Q_{3}$), precision, recall, specificity and MCC. The mathematical definition of these measures is found through equations \ref{eq:accuracy}, \ref{eq:precison}, \ref{eq:recall}, \ref{eq:specificity} and \ref{eq:MCC}[]. 
\begin{equation}
\label{eq:accuracy}
\;\;\;\;\;\;\;\;\;\;\;\ Overal Accuracy = \frac{ \textstyle \sum_{j} TP_{j}+TN_{j}}{ \textstyle \sum_{j} TP_{j}+TN_{j}+FP_{j}+FN_{j}}
\end{equation}
\vspace{-2mm}
\begin{equation}
\label{eq:precison}
\;\;\;\;\;\;\;\;\;\;\;\ Precision_{j} = \frac{TP_{j}}{TP_{j}+FP_{j}}
\end{equation}
\vspace{-2mm}
\begin{equation}
\label{eq:recall}
\;\;\;\;\;\;\;\;\;\;\;\ Recall_{j} = \frac{TP_{j}}{TP_{j}+FN_{j}}
\end{equation}
\vspace{-2mm}
\begin{equation}
\label{eq:specificity}
\;\;\;\;\;\;\;\;\;\;\;\ Specificity_{j} = \frac{TN_{j}}{TN_{j}+FP_{j}} 
\end{equation}
\vspace{-2mm}
\begin{equation}
\label{eq:MCC}
\;\;\;\;\;\;\;\;\;\;\;\;\;\;\ MCC_{j} = \frac{TP_{j}.TN_{j}-FP_{j}.FN_{j}}{\sqrt{(TP_{j}+ FP_{j})(TP_{j} + FN_{j})(TN_{j} + FP_{j})(TN_{j} + FN_{j})}}
\end{equation}
Where TP, FP, TN and FN respectively stand for true positive, false positive, true negative and false negative from the confusion matrix.
\vspace{-6mm}
\subsection{Availability}
\label{sec:Availability}
\vspace{-4mm}
The links to download our code and the tree data sets are available at: 
\vspace{-8mm}
\section{methodology}
\label{subsec:method}
\vspace{-4mm}
In this section, we explain the components of our framework as shown in Fig. \ref{fig:frameOver}.
\vspace{-11mm}
\subsection{Pre-processing}
\label{sec:preProcc}
\vspace{-4mm}
Conventional machine learning-based methods for prediction of protein secondary structure generate numerical feature vectors from the sequences of protein primary structure. The feature vectors are subsequently fed into a learning machine for structure prediction. Although such numerical features comprise protein evolutionary information or biochemical properties, they cause information loss in comparison with the original sequence. Additionally, some numerical encoding schemes can cause high dimensional feature vectors. In fact, the learner's performance can differ based on various collection of numerical features. All the aforementioned consequences of numerical feature extraction can affect the learner's performance negatively. As a result, in this work, we bypass feature extraction and directly process the input protein sequences. Thus the only pre-processing procedure we perform is to applying a sliding window of size $h=17$ on sequences of protein primary structure varying in length. 
The neighbors of a residue in a protein sequence contribute strongly to its secondary structure. Therefore, the sliding window technique facilitates the involvement of a residue's neighbors in structure prediction. We predict the secondary structure of each residue at the center of the window that involves $(h-1)/2$ neighbors on the left and right sides. Assuming that there are $M$ residues in the protein dataset, sliding the window will lead to $M$ sequences (i.e. $P_{i}^h$) where each $P_{i}^h$ locates one of the data set residues at the center. We use the windows size that is justified by \cite{Zangooei2012,Spencer2015}. Excessively large $h$ deviates the dissimilarity values between sequence pairs. Furthermore, the value of $h$ can not be very small (i.e. $h<5$) because the dissimilarity measure includes computation of n-gram scores and extremely small $n$ is biologically meaningless. Logically the value for $n$ must be smaller than $h$. 
\vspace{-6mm}
\subsection{The compound Dissimilarity Measure}
\label{sec:preProcc}
\vspace{-4mm}
The $\widetilde{d}$-FKNN component performs classification based on the compound dissimilarity measure $\widetilde{d}$ that is computed using three parameters including LZ-scores, n-gram scores, and the $\rho_d$ dissimilarity rate. Our experiments show that the addition of each parameter to $\widetilde{ d }$ can further improve the accuracy (Section \ref{sec:evaluation}).\\
The LZ complexity measure reflects the degree of repeated patterns or the level of randomness in a sequence and can include the position information \cite{Tan2015,Liu2010}.\\ 
Assume $p_{i}^h[m:n]$ is a fragment of a protein sequence starting at position $m$ and ending at position $n$ where $1<m<n<h$. Thus we can show that $p_{i}^h = p_{i}^h[1:m_{1}]p_{i}^h[m_{1}+1:m_{2}]...p_{i}^h[m_{k-1}+1:m_{k}]...p_{i}^h[m_{z-1}+1:h]$. The LZ complexity of the protein sequence $c(p_{i}^h)$ is the number of fragments in the exhaustive history that represents the decomposition of the sequence. Each fragment obtained from the decomposition process must be unique except in the last step, at which it is permitted to copy a previously generated fragment.\\
For example, for the protein sequence $p_{i}^h = TTCCPSTCIVPSA$ the exhaustive history is $T.TC.C.P.S.TCI.V.PSA$ where '.' separates the fragments generated at each step. Hence $c(p_{i}^h$) is 8 which is the number of fragments in the exhaustive history. The initial LZ score between two protein sequences ($ \zeta( p_{i}^h, p_{j}^h ) $) is defined in Eq.1 \cite{Liu2010}.
\vspace{-6mm}
\begin{equation}
\;\;\;\;\;\;\;\;\;\;\;\;\;\;\;\;\;\;\;\;\;\zeta( p_{i}^h, p_{j}^h ) =  c( p_{i}^hp_{j}^h) - c( p_{i}^h)
\end{equation}
Where $p_{i}^hp_{j}^h$ is the concatenation of two protein sequences of $p_{i}^h$ and $p_{j}^h$. The more similar the two sequences are, the less the vale of $\zeta(.)$ will be. The final LZ score (the LZ dissimilarity of two sequences) is attained from the normalization of Eq.1 as formulated in Eq.2. We utilize the normalized LZ score in our work \cite{Liu2010}.
\vspace{-4mm}
\begin{equation}
\;\;\;\;\;\;\;\;\;\;\;\;\;\;\;\;\;\;\;\;\;LZ = \frac{max\left\{ \zeta\left( p_{i}^h, p_{j}^h \right) , \zeta\left( p_{j}^h, p_{i}h \right)\right\}}{max\left\{ c\left( p_{i}^h \right) , c\left( p_{j}^h \right)\right\} }
\end{equation}\\
Since each string has a unique exhaustive history \cite{Lempel1976}, algorithm 1 demonstrates how we generate the exhaustive history of a sequence.\\
\vspace{-5mm}
\begin{algorithm}[H]
	\SetAlgoLined
	\SetKw{KwGoTo}{go to}
	\text{}\\
	\label{alg1}
	\textbf{Input:} $p_{i}^h$ (a protein sequence with length h).\\
	\textbf{Output:} $\eta(p_{i}^h)$ (set of fragments related to the exhaustive history of protein sequence).
	\begin{algorithmic}[1]
		\STATE $\eta(p_{i}^h)\leftarrow\emptyset,\: i=1,\: j=1 $\
		\STATE $\eta(p_{i}^h) \leftarrow p_{i}^h[i],\: i+=1,\: j+=1$\
		\IF{$j >= h$} {\label{h}}
		\STATE $\eta(p_{i}^h)\leftarrow p_{i}^h[i:j]$ and terminate\
		\ENDIF
		\IF{$p_{i}^h[i:j]\notin\eta(p_{i}^h)$}
		\STATE  $\eta(p_{i}^h) \leftarrow p_{i}^h[i:j],  i = j+1 ,  j += 1$, \KwGoTo \ref{h}\ 
		\ELSE 
		\STATE $j += 1$, \KwGoTo \ref{h}
		\ENDIF
		\caption{create the exhaustive history of a protein sequence}
	\end{algorithmic}
\end{algorithm}
\clearpage
As stated previously, LZ complexity for a sequence considers the character distribution rather than the characters themselves. Hence, this property leads to the same complexity for sequences of the same distributions, but with different characters. As a case in point, consider the two sequences of $p_{i}^h$=$APAFSVSGG$ and $p_{j}^h$=$THTDKRKLL$. The exhaustive history of the sequences are $\eta(p_{i}^h)=A.P.AF.S.V.SG.G$ and $\eta(p_{j}^h)$=$T.H.TD.K.R.KL.L$. However the LZ complexity for both strings is identical and equal to 7. Nevertheless, each amino-acid brings along distinct properties to form a certain secondary structure. Therefore, to include sensitivity to the type of amino-acid molecules, we employ a new parameter called dissimilarity rate $\rho_{d}$. Let $\pi_{i}^m$ and $\pi_{j}^m$ be the respective list of unique amin-acids composing $p_{i}^h$ and $p_{j}^h$ while $m_{i}\in\pi_{i}^m$ and  $m_{j}\in\pi_{j}^m$. Here, $\rho_{d}$ is retrieved by substituting of Eq.4 and Eq.5 into Eq.3.
\begin{gather}
\rho_{d}=\frac{1+\vert\delta_{i,j}\vert}{1+\vert\mu_{i,j}\vert}\\
\delta_{i,j}=\lbrace m_{i} \cup m_{j} \vert m_{i}\in \pi^{m}_{i} and\:\: m_{j}\in \pi^{m}_{j}\rbrace \backslash \lbrace m_{i} \vert m_{i}\in \pi^{m}_{i} and\:\: m_{i}\in \pi^{m}_{j}\rbrace\\
\mu_{i,j} = \lbrace m_{i} \vert m_{i}\in \pi^{m}_{i} and\:\: m_{i}\in \pi^{m}_{j}\rbrace
\end{gather}
As a significant parameter for secondary structure prediction, Local similarities among protein sequences can identify conserved structures during proteins' evolution \cite{Lin2010}. To incorporate the local similarities into our measure, we employ n-gram score $\vert\Gamma{i,j}^n\vert$ between each pair of sequence ($p_{i}^h, p_{j}^h$). The larger the n, the more strictly the similarity will be computed. Let $\Gamma_{i}^n$ and $\Gamma_{j}^n$ be the respective sets of n-gram patterns associated with the protein sequences of $p_{i}^h$ and $p_{j}^h$ where $\gamma_{i}^n\in \Gamma_{i}^n$ and $\gamma_{j}^n\in \Gamma_{j}^n$. We compute the n-gram score ($ \Gamma_{i,j}^n $) using Eq.6.
\vspace{-2mm} 
\begin{gather}
\Gamma_{i,j}^n=\lbrace \gamma_{i}^n\vert \gamma_{i}^n\in \Gamma_{i}^n\: and\: \gamma_{i}^n\in \Gamma_{j}^n\rbrace 
\end{gather}
Algorithm 2 shows how in this paper, we obtain the n-gram patterns of a sequence.\\
\begin{algorithm}[H]
	\SetAlgoLined
	\SetKw{KwGoTo}{go to}
	\text{}\\
	\label{alg2}
	\textbf{Input:} $p_{i}^h$ (a protein sequence with length $h$).\\
	\textbf{Output:} $\Gamma_{i,j}^n$ (the set of n-gram patterns ($n<h$) related to the input protein sequence).
	\begin{algorithmic}[1]
		\STATE $\Gamma_{i,j}^n\leftarrow\emptyset,\: i=0$\
		\WHILE {$i+n\leq h$}
		\STATE $\Gamma_{i,j}^n\leftarrow p_{i}^h[i+1:i+n]$, $i+=1$
		\ENDWHILE
		\caption{generate the n-gram patterns of a protein sequence}
	\end{algorithmic}
\end{algorithm}
\text{}\\
We fuse LZ scores, n-gram scores ($\vert\Gamma{i,j}^n\vert$) and the dissimilarity rate ($\rho_{d}$) into Eq.7 to create 
our final dissimilarity measure (i.e. $\widetilde{ d }$). As $\widetilde{ d }$ is a compound measure with beneficial parameters, it can infer dissimilarity more effectively.
\vspace{-1mm}
\begin{equation}
\tilde{ d } = \frac{max\left\{ \zeta\left( p_{i}^h, p_{j}^h \right) , \zeta\left( p_{j}^h, p_{i}h \right)\right\} \times\left( 1 + \vert\mu_{i,j}\vert \right)}{max\left\{ c\left( p_{i}^h \right) , c\left( p_{j}^h \right)\right\} \times\left( 1 + \vert\delta_{i,j}\vert \right) \times \left( 1 + \vert\Gamma_{i,j}^n\vert\right)}
\end{equation}\\
\vspace{-12mm}
\subsection{Fuzzy KNN Algorithm}
\label{sec:Fuzz_Knn}
\vspace{-4mm}
unlike some model-based learners such as rule-based methods and decision trees whose boundaries are triangular and straight lines \cite{Tan2006}, the KNN is capable to implement the complicated and irregular decision boundaries. Nevertheless, the effectiveness of KNN method \cite{Han2011,Theodoridis2003} is significantly influenced by the selected distance measure. In this paper, we introduced the effective dissimilarity measure of $\widetilde{ d }$. We later employ $\widetilde{ d }$ in the membership function ($U_{c}(.)$) of the fuzzy KNN ($\widetilde{ d }$-FKNN) to determine the nearest neighbors of the input data. Precisely speaking, FKNN does not return one secondary structure for each input residue ($p_{i}^h[r]$). Rather, it produces the likelihood for the input residue to adopt each secondary structure. Eq.8 \cite{Keller1985} computes the fuzzy membership values of a test residue ($p^h_{i}[r]$) in each secondary class.
\vspace{-3mm}
\begin{equation}
\small
\label{eq:Framework}
U_{c}(p^h_{i}[r])=\frac{\sum_{j=1}^{k}I_{cj}\left( \frac{1}{\left\Vert p^h_{i}-p^h_{j} \right\Vert^{\frac{2}{m-1}}} \right)}{\sum_{j=1}^{k}\left( \frac{1}{\left\Vert p^h_{i}-p^h_{j} \right\Vert^{\frac{2}{m-1}}} \right)}
\end{equation}
Here, $U_{c}(p^h_{i}[r])$ is the likelihood of the unlabeled input residue $p^h_{i}[r]$ to belong to class $c$. Also, $k$ is the number of neighbors, $\Vert . \Vert$ denotes the degree of dissimilarity between $p^h_{i}$ and its neighbor ($p^h_{j}$) and $m$ determines the degree of fuzziness. Finally, $ U_{cj}$ is the initial fuzzy membership value of the neighbor $p^h_{j}$ for class $c$. As formulated in Eq.8, the likelihood for each test residue ($p^h_{i}[r]$) belonging to each of three classes is dependent on the initial membership values of the neighbors and also the inverse dissimilarity between $p^h_{i}[r]$ and its neighbors ($p^h_{j}[r]$). The inverse dissimilarity ($ \frac{1}{\left\Vert p^h_{i}-p^h_{j} \right\Vert} $) determines how each neighbor can enforce the membership of the input residue in a class (c). In fact, as far as a neighbor is from the input residue, in a decaying manner less weight will be assigned to the neighbor [42]. Eq.9 \cite{Keller1985} indicates how we compute the initial membership values corresponding to the neighbors ($ I_{c}(p^h_{j})$). Suppose the class of the training residue with known structure ($ p^h_{j} $) is $s$.
\vspace{-2mm}
\begin{equation}
\small
\label{eq:Framework}
I_{c}(p^h_{j}[r])= 
\begin{cases}
0.51+(\frac{n_{j}}{\acute{k}})\times0.49 & \text{for}\: c = s\\
(\frac{n_{j}}{\acute{k}})\times0.49 & \text{for}\: c\neq s
\end{cases}
\end{equation}
Where $I_{c}(p^h_{j}[r])$ is the corresponding membership value of a training residue ($ p^h_{j}[r] $) for class $c$, $\acute{k}$ is the number of neighboring training residues of $ p^h_{j} $ and $n_{j}$ is the number of neighboring training residues of $  p^h_{j} $ that also belong to class $s$. $I_{c}(.)$ assigns three initial values of fuzzy membership ($ c \in \lbrace H, E, C\rbrace $) to each training residue $p^h_{i}[r]$ which corresponds to each class of secondary structure. According to Eq.9, if the label of $p^h_{j}[r]$ and its neighbors is $c$, $I_{c}(p^h_{j}[r])$ gets the full membership of 1 for class $c$. However, if $p^h_{j}[r] \neq c$ or its neighbors do not belong to class $c$, $p^h_{j}[r]$ gets a membership value of less than 1 for class $c$ . In other words, the function $ I_{c}(.) $ aims to fuzzify the class membership of the labeled residues (with known structure) which lie in the intersecting region of three classes in the sample space. However, $ I_{c}(.) $ assigns full membership value of 1 to the samples far away from the intersecting region. As the value of $ U_{c}(p^h_{i}[r]) $ is computed using $I_{c}(p^h_{j}[r])$ (where $p^h_{j}$ is the neighbor of $p^h_{i}$), the class of unlabeled residues ($ p^h_{i}[r] $) located in the intersecting region of classes will be less influenced by the labeled residues ($ p^h_{j}[r] $) lying in the intersecting region \cite{Keller1985}. Assignment of the initial fuzzy membership values to the training residues in fuzzy KNN can be considered a training phase, which leads to performance enhancement compared to the crisp KNN. Another important advantage of the employed fuzzy property is that the fuzzy membership values provide a confidence level for predictions which determines how strongly an input data belongs to each class. 
\vspace{-7mm}
\subsection{Edit-SVM Algorithm}
\label{sec:edit_SVM}
\vspace{-4mm}
The power and efficiency of SVM lie in the fact that it actually solves an optimization problem to find the globally optimum solution. The SVM not only finds the separating boundary, but also discovers the boundary with the maximum margin from the samples of each class. This property boosts the generalization ability of SVM. Furthermore, SVM searches a higher dimensional feature space in order to find a separating hyperplane rather than searching the original space to find a non-linear separating boundary \cite{Han2011}. Using kernel tricks, the transformation to higher dimensions does not need to be directly calculated and thus the efficiency of the algorithm remains high. This algorithm outperforms the shallow neural networks and almost all individual learners for prediction of protein secondary structure \cite{Bouziane2015,Zangooei2012a}. Thus, it has been selected as one component of our framework.
In the SVM module, we aim to directly process the protein language and predict the structure of the residue that is positioned at the center of the sliding window. Therefore, we employ a kernel capable to take string data as input. A popular and well-performing kernel in this area is the \textit{edit} kernel, formulated in Eq.10 \cite{Aygun2010}, which utilizes the \textit{edit} distance between two strings in a form similar to RBF kernel. RBF kernel is one of the most effective kernels that maps the original feature space to an infinite dimensional space \cite{Zangooei2012a,Theodoridis2003}. 
\vspace{-2mm}
\begin{equation}
\small
\label{eq:Framework}
K(x,y)=e^{\gamma\cdot edit(x,y)}
\end{equation}
Given two strings $x$ and $y$ generated from the alphabet set $\Sigma$, the \textit{edit} distance ($edit(x,y)$) is the minimum number of operations that transform $x$ to $y$. \textit{Levenshtein} ia an \textit{edit} distance that permits the operations of insertion, deletion, and substitution \cite{VanderLoo2014} when transforming strings.
\vspace{-7mm}
\subsection{Biological Filtering}
\label{sec:filter}
\vspace{-4mm}
Considering a protein sequence (i.e. $p_{i}^h$) composed of $n$ amino-acids and three classes of secondary structure, the prediction vector will have $3^n$ different states. Nevertheless, not all of these states are biologically meaningful and valid. For instance, a single $H$ structure in a structure sequence is meaningless. Hence, in order to rectify the prediction output, we apply a set of biological transformations proposed by \cite{Alirezaee2012,Bondugula2005} on the secondary structure sequences. The transformers are elucidated as follows:
${EHE}\rightarrow{EEE}$,	${HEH}\rightarrow{HHH}$,	${HCH}\rightarrow{HHH}$,	${ECE}\rightarrow{EEE}$ and ${HEEH}\rightarrow{HHHH}$. \\
For an instance ${EHE}\rightarrow{EEE}$ means that if the chunk $ EHE $ is observed in a protein sequence, it will be transformed to $ EEE $. Because a single $H$ structure cannot appear between two \textit{E} structures. It is important to remind that applying such filtration on the output of different predictors can cause varying results. In some methods, enacting filtration on the prediction vectors can significantly improve the accuracy. Such methods are more accurate since they have more single false predictions rather than contiguous mispredictions. However, some methods may show a minor improvement over filtration.
\vspace{-6mm}
\subsection{Aggregation Rules}
\label{sec:AggRul}
\begin{figure}[H]
	\centering
	\includegraphics[trim=0cm 0cm 0cm 0cm,clip=true,scale=0.45]{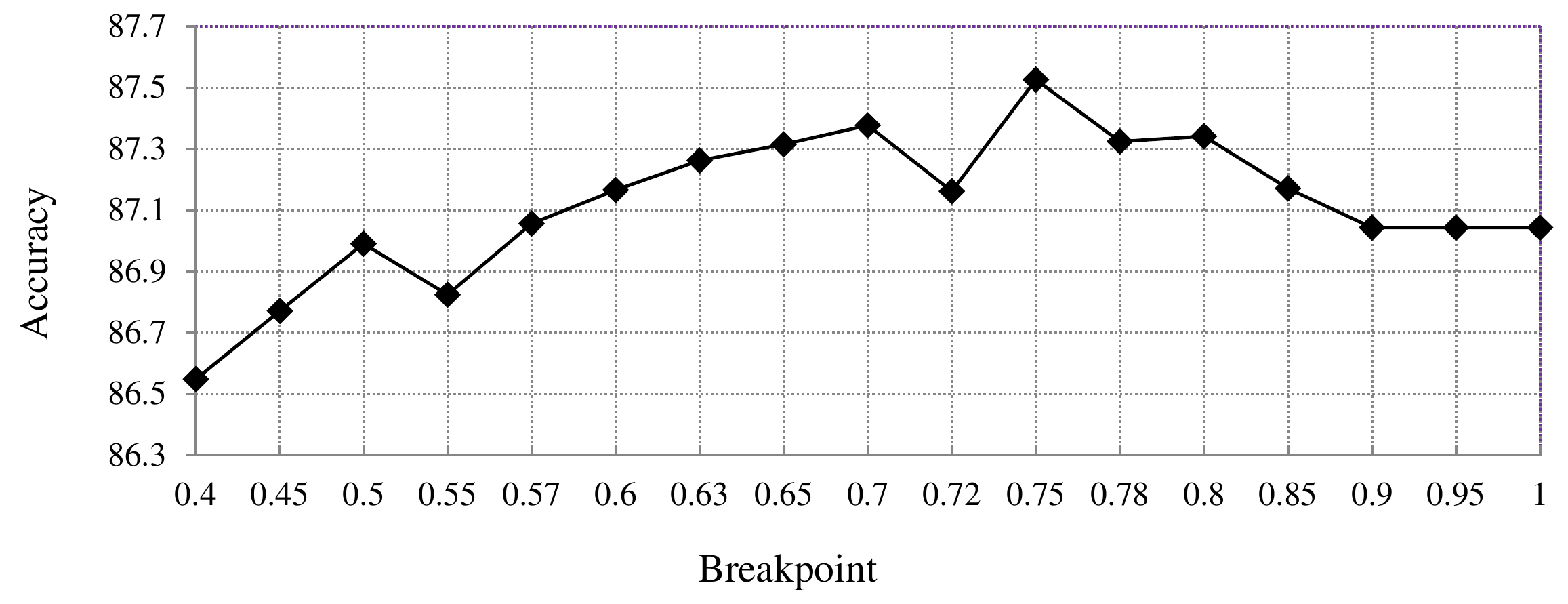}
	\vspace{-3mm}
	\caption{Impact of interval breakpoints on accuracy}
	\label{fig:intervalBreak}
\end{figure}

\vspace{-4mm}
Aggregation is an influential module in the effectiveness of a multi-component learner \cite{Bouziane2015}. In our work, five various aggregation rules are proposed to produce the final secondary structures. The aggregators take advantage of the output from  $\widetilde{ d }$-FKNN and edit-SVM classifiers. $\widetilde{ d }$-FKNN generates three fuzzy membership values which provide a confidence level for each decided class and enhance the aggregation process. The accuracy of our method using each aggregation rule is evaluated in section \ref{sec:experimental}.\\
Let $p_{i}^h[r]$ be the $r^{th}$ residue of the protein sequence $p_{i}^h$. Also, assume $\Delta_{FKNN}^1$ and $\Delta_{FKNN}^2$ to be the first and second decisions (associated with the class with maximum and mid membership values) made by $\widetilde{ d }$-FKNN classifier. Suppose $s_{i}[r]$ to be the predicted secondary structure for the residue $p_{i}^h[r]$.\\
Aggregation 1: $\forall\: p_{i}^h[r],\: if\:\: \lbrace\Delta_{FKNN}^1=\Delta_{SVM}\rbrace\:\: then\\
\newcommand\tab[1][1cm]{\hspace*{#1}}
\tab\tab\tab\tab\tab s_{i}[r]=\Delta_{FKNN}^1\\
\tab\tab\tab\tab else\\
\tab\tab\tab\tab\tab s_{i}[r]=\Delta_{FKNN}^2\\
\tab\tab\tab\tab endif$\\
According to aggregation 1, if both $\widetilde{ d }$-FKNN and edit-SVM vote to a certain secondary structure, it will be taken as the final prediction. Otherwise, based on the fuzzy levels of confidence, the second decision of $\widetilde{ d }$-FKNN ($ \Delta_{FKNN}^2 $) will most probably be the actual prediction outcome. Aggregations 2, 4 and 5 work based on the weighted decisions of the classifiers. We introduce two strategies for weight assignment. The first strategy assigns a weight proportional to the accuracy of the classifier on a validation set. Accordingly, the more accurate the classifier is, the higher priority its decision will get in the aggregation process. Suppose $ \omega_{1} $ and $ \omega_{2}$ are the weights for either of $\widetilde{d}$-FKNN and edit-SVM where $ \omega_{1}\leq\omega_{2}$ and $ \omega_{1}+\omega_{2}=1$. Eq.10 and 11 are used to compute the weights.
\vspace{-2mm}
\begin{equation}
\omega_{1}=\frac{min\left\{ \alpha_{fknn}, \alpha_{svm} \right\}}{\alpha_{fknn}+ \alpha_{svm}}
\end{equation}
\vspace{-6mm}
\begin{equation}
\omega_{2}=1-\omega_{1}
\end{equation}
Subsequently, the interval of [0,1] is divided into two sub-intervals proportional to the weight of each classifier. For instance, if $\omega_{1}=0.6$ and $\omega_{2}=0.4$, the sub-intervals of $i_{1}=[0-0.6]$ and $i_{2}=(0.6-1]$ will be dedicated to the first and the second classifiers respectively. Finally, we generate a random number $r$ in the interval of [0,1]. If the value of $r$ lies in $i_{1}$, the first classifier will decide the final class. Otherwise, the second classifier will determine the final prediction. In fact, the length of the dedicated sub-interval to a certain classifier is the probability that the decision of that classifier is returned as the final decision. We call this strategy as \textit{Roulette Wheel 1} which is particularly more applicable when the accuracy of the classifiers are not level.\\
The second strategy (\textit{Roulette Wheel 2}) differs from the \textit{Roulette Wheel 1} when dedicating a sub-interval to each classifier. Instead of assigning sub-intervals according to the predefined weights of the classifiers, \textit{Roulette Wheel 2} considers a step size and examines the resultant accuracy of the prediction machine. Here we appoint different breakpoints over the interval of [0,1] and the selected breakpoint assigns a sub-interval to each classifier. For example, if the step size is set to 0.1, the breakpoints $\lbrace 0.1, 0.2, 0.3,..,0.9\rbrace$ will be examined and the sub-intervals of the classifiers can be appointed in pairs as $i_{1}=[0,0.1]$ and $i_{2}=(0.1,1]$, $i_{1}=[0,0.2]$ and $i_{2}(0.2,1]$, $i_{1}=[0,0.3]$ and $i_{2}(0.3,1]$,..., and $i_{1}=[0,0.9]$ and $i_{2}=(0.9,1]$.\\
In this work, the accuracy of the two classifiers is nearly even. Hence we employ the weight assignment strategy of \textit{Roulette Wheel 2} in aggregation rules of 2, 4 and 5. Figure \ref{fig:intervalBreak} illustrates the change in accuracy (of the prediction machine) when an increase in the value of breakpoint is observed.
\vspace{-4mm}
\vspace{-3mm}
In this experiment, we compute the accuracy as an average of 15 generated random numbers (r) for each breakpoint. As shown in Figure \ref{fig:intervalBreak}, the breakpoint 0.75 which assigns the interval of [0,75) to $\widetilde{ d }$-FKNN and [0.75,1] to edit-SVM gains to the highest accuracy.\\
The pseudo-code of aggregation 2 is described below:\\
Aggregation 2: $\forall\: p_{i}^h[r],\: if\:\: {\Delta_{FKNN}^1=\Delta_{SVM}}\:\: then\\
\newcommand\tab[1][1cm]{\hspace*{#1}}
\tab\tab\tab\tab\tab s_{i}[r]=\Delta_{FKNN}^1\\
\tab\tab\tab\tab else\\
\tab\tab\tab\tab\tab if\: r\leq\omega_{FKNN}\\
\tab\tab\tab\tab\tab\tab s_{i}[r]=\Delta_{FKNN}^1\\
\tab\tab\tab\tab\tab else \\
\tab\tab\tab\tab\tab\tab s_{i}[r]=\Delta_{SVM}\\
\tab\tab\tab\tab\tab endif \\
\tab\tab\tab\tab endif$\\
According to aggregation 2, where $\Delta_{FKNN}^1$ and $\Delta_{SVM}$ do not equate, the \textit{Roulette Wheel 2} is performed to determine the final prediction from $\Delta_{FKNN}^1$ and $\Delta_{SVM}$ based on the associated weights.\\
Aggregation 3: $\forall\: p_{i}^h[r],\: if\:\: {\Delta_{FKNN}^1=\Delta_{SVM}}\:\: then\\
\newcommand\tab[1][1cm]{\hspace*{#1}}
\tab\tab\tab\tab\tab s_{i}[r]=\Delta_{FKNN}^1\\
\tab\tab\tab\tab else\\
\tab\tab\tab\tab\tab s_{i}[r]=\Delta_{FKNN}^3\\
\tab\tab\tab\tab endif$\\
As aggregation 3 denotes, when $\Delta_{FKNN}^1$ differs from $\Delta_{SVM}$, the last fuzzy decision of KNN (i.e. $\Delta_{FKNN}^3$) will predict the final structure. Intuitively, selecting the last fuzzy decision is not necessarily the best choice. Nevertheless, we opt for the last fuzzy decision as we aim to investigate how the fuzzy decisions comply with both edit-SVM decision and the sequence-structure relation.\\
Aggregation 4: $\forall\: p_{i}^h[r],\: if\:\: {\Delta_{FKNN}^1=\Delta_{SVM}}\:\: then\\
\newcommand\tab[1][1cm]{\hspace*{#1}}
\tab\tab\tab\tab\tab s_{i}[r]=\Delta_{FKNN}^1\\
\tab\tab\tab\tab else\\
\tab\tab\tab\tab\tab if\: r\leq\omega^2_{FKNN}\\
\tab\tab\tab\tab\tab\tab s_{i}[r]=\Delta_{FKNN}^2\\
\tab\tab\tab\tab\tab else \\
\tab\tab\tab\tab\tab\tab s_{i}[r]=\Delta_{FKNN}^3\\
\tab\tab\tab\tab\tab endif \\
\tab\tab\tab\tab endif$\\
In aggregation 4, if the two classifiers predict different structures, the \textit{Roulette Wheel 2} will use $\Delta_{FKNN}^2$ and $\Delta_{FKNN}^3$ to select the final structure. Aggregation 4 states that an improper local decision can cause an improper global decision.\\
Aggregation 5: $\forall\: p_{i}^h[r],\: if\:\: {\Delta_{FKNN}^1=\Delta_{SVM}}\:\: then\\
\newcommand\tab[1][1cm]{\hspace*{#1}}
\tab\tab\tab\tab\tab s_{i}[r]=\Delta_{FKNN}^1\\
\tab\tab\tab\tab else\\
\tab\tab\tab\tab\tab if\: r\leq\omega_{FKNN}\\
\tab\tab\tab\tab\tab\tab s_{i}[r]=\beta(\Delta_{FKNN}^1)\\
\tab\tab\tab\tab\tab else \\
\tab\tab\tab\tab\tab\tab s_{i}[r]=\beta(\Delta_{SVM})\\
\tab\tab\tab\tab\tab endif \\
\tab\tab\tab\tab endif$\\
Aggregations 5 and 2 work quite similarly except that, aggregation 5 receives the filtered output from $\widetilde{ d }$-FKNN ($\beta(\Delta_{FKNN}$)) and edit-SVM ($\beta(\Delta_{SVM}$)). Theoretically, aggregation 2 and 5 can provide better results compared to other rules. Because they rely on the early fuzzy decisions.\\
\vspace{-5mm}
\begin{proof} {\textbf{\textit{Performing classification using MCP on an arbitrary number of classes}}}
\label{sec:extendedAgg}
Considering the number of classes to be an arbitrary value of $l$, there will be a unified final decision corresponding to SVM ($\Delta_{SVM}$) which is obtained via voting between the decisions of $\frac{l(l-1)}{2}$ SVM models (one-versus-one strategy \cite{Ward2003}). There also will be $l$ decisions associated with FKNN as $\Delta_{FKNN}^1$, $\Delta_{FKNN}^2$, ..., $\Delta_{FKNN}^l$  where $\Delta_{FKNN}^1$ and $\Delta_{FKNN}^l$ correspond to the secondary class with maximum and minimum values of fuzzy membership function respectively. In aggregation rules of 1, 2, and 5, we take advantage of the decision of SVM ($\Delta_{SVM}$) and the first decision of FKNN ($\Delta_{FKNN}^1$). Thus, MCP can perform classification on an arbitrary number of classes. 
\end{proof} 
\vspace{-2mm}
\begin{table}[H]
	\footnotesize
	\centering
	\caption{The accuracy of the propped approach and its components} 
	\label{table:perocessDev} 
	\begin{tabular}{p{15em}:l}
		\toprule
		\textbf{Method} & \multicolumn{1}{p{6em}}{\textbf{Q3}} \\
		\midrule
		FKNN+LZ & 76.38 \\[4pt]
		FKNN+LZ+$\rho_d$ & 80.42 \\[4pt]
		FKNN+LZ+$\rho_d$+n-gram ($\widetilde{d}$) & 81.96 \\ [4pt]    
		Edit-SVM & 82.2701 \\[4pt]
		MCP1 & 83.0906 \\ [4pt]
		MCP2 & \textbf{85.4072} \\[4pt]
		MCP3 & 75.487 \\[4pt]
		MCP4 & 80.8661 \\[4pt]
		MCP5 & \textbf{87.2853} \\[4pt]
		\bottomrule
	\end{tabular}%
	\label{tab:addlabel}%
	\vspace{-2mm}
\end{table}
\vspace{-9mm}
\section{Experiment}
\label{sec:experimental}
\vspace{-3mm}
We conducted comprehensive experiments on real-world RS126 benchmark dataset to evaluate the effectiveness of our framework in the prediction of the protein secondary structure. Additionally, we compared our approach with four state-of-the-art models to validate the competence of our proposed framework. We also consider four widely-used evaluation measures of accuracy, recall, specificity, and Matthews Correlation Coefficient (MCC) to compare the classification baselines.
At the proper time during the course of development, we used $Q_{3}$ accuracy to gradually evaluate the performance of our proposed framework in protein structure prediction. Table \ref{table:perocessDev} reports the comparison among the effectiveness of various aggregation rules. From the results in Table \ref{table:perocessDev}, we can see that adjoining of each component (e.g.  the dissimilarity parameter) to the primary ensemble machine can enhance the overall performance of the framework. According to Table \ref{table:perocessDev}, extending the FKNN model with the dissimilarity rate ($ \rho_d $) significantly improves the accuracy. It is also observed that the multi-component variations (i.e. MCP1 to MCP5) remarkably perform better than the single classifiers (e.g. $\widetilde{ d }$-FKNN and edit-SVM). Our framework gains a better effectiveness compared to other competitors. The reason is that our framework is not only multi-component, but also takes advantage of a new dissimilarity measure alongside with different aggregation rules. Due to utilizing the first fuzzy and edit-SVM decisions and including an extra filtration, aggregation rules of 1, 2 and 5 perform better than other variations. Because of granting priority to the last and second fuzzy decisions, both aggregation rules of 3 and 4 gain lower accuracy . Consequently, we exclude the aggregation rules of 3 and 4 from the rest of the experiments.
\vspace{-4mm}
\begin{figure}[H]
	\centering
	\minipage{0.35\textwidth}
	\centering
	\label{fig:a}
	\includegraphics[width=\linewidth]{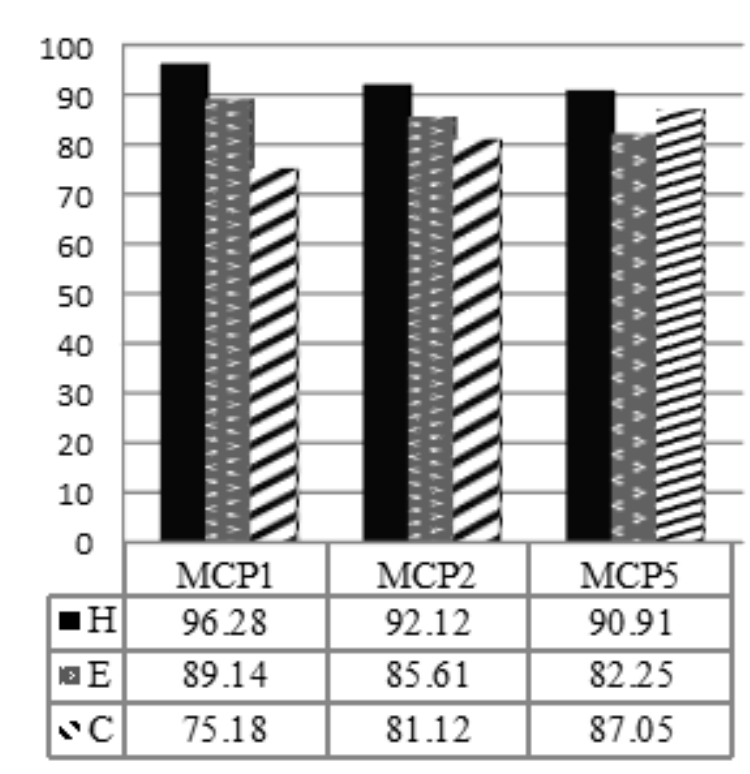} 
	\small (a) Precision	
	\endminipage\hfill	
	\minipage{0.35\textwidth}
	\centering
	\label{fig:b}
	\includegraphics[width=\linewidth]{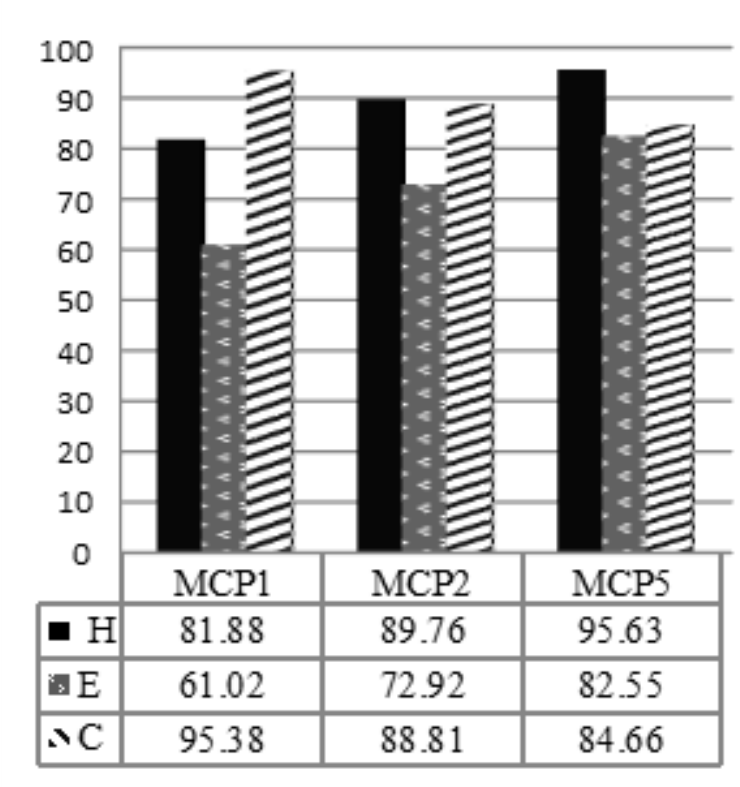} 
	\small (b) Recall	
	\endminipage\hfill
	\minipage{0.35\textwidth}
	\centering
	\label{fig:c}
	\includegraphics[width=\linewidth]{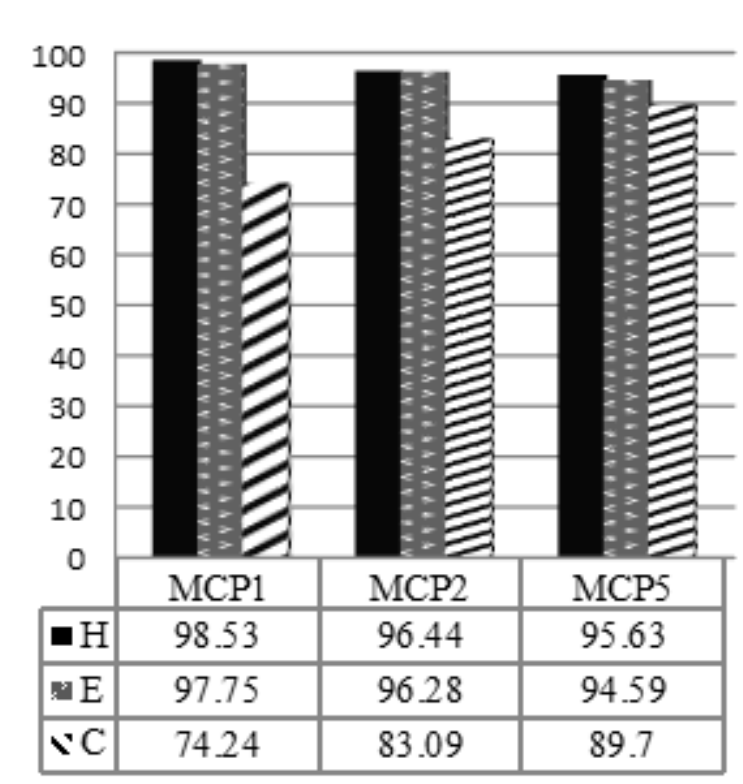} 
	\small (c) Specificity
	
	\endminipage\hfill
	\minipage{0.35\textwidth}
	\centering
	\label{fig:d}
	\includegraphics[width=\linewidth]{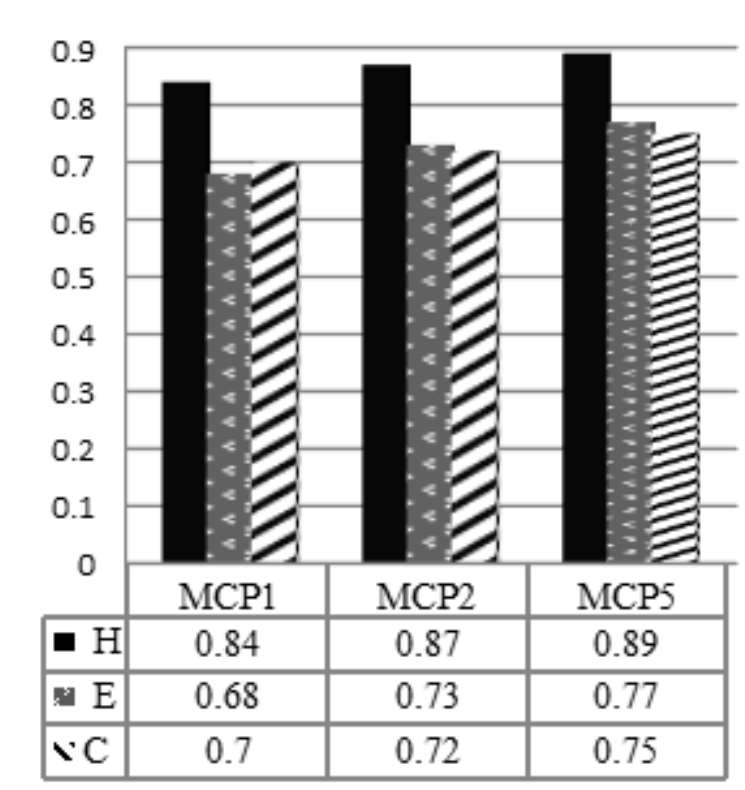} 
	\small (d) MCC
	
	\endminipage				
	\vspace{-2mm}
	\caption{Effectiveness comparison between MCP1, MCP2, and MCP5}
	\label{fig:allMeasAlllClass}
\end{figure}
\vspace{-5mm}
Figure \ref{fig:allMeasAlllClass} compares the effectiveness of various version of our proposed framework using precision, recall, specificity, and MCC metrics. According to Figure \ref{fig:allMeasAlllClass}(a), MCP1 has the highest precision in predicting \textit{H} and \textit{E} structures. However, it shows a fairly low precision for the class \textit{C}. Also, compared to MCP1, the MCP2 model demonstrates a better precision on class \textit{C}. Moreover, since the transformations of ${HCH}\rightarrow{HHH}$ and ${ECE}\rightarrow{EEE}$ reduce the number of false-positives, the extra filtration in MCP5 significantly promotes the precision for the \textit{C} structures. Despite the fact that MCP5 gains a lower precision for \textit{H} and \textit{E} structures as shown in Figure \ref{fig:allMeasAlllClass}(b), MCP5 notably shows a better recall compared to MCP1 and MCP2 in the prediction of \textit{H} and \textit{E} structures.\\
As depicted in Figure \ref{fig:allMeasAlllClass}(b), MCP1 is sensitive to the class imbalance. Since the higher portion of structures respectively belongs to class \textit{C}, \textit{H}, and \textit{E}, MCP1 tends to predict each sample as \textit{C}, \textit{H}, and \textit{E} respectively. This imbalanced tendency leads to lower false negative predictions for each of the classes based on the number of data samples. Hence, MCP1 exhibits the maximum and minimum recall values for the classes of \textit{C} and \textit{E} respectively. Nevertheless, while MCP2 partially resolves this sensitivity, MCP5 remarkably eliminates the effect of the class imbalance.\\
As illustrated in Figure \ref{fig:allMeasAlllClass}(c), all the variations of our model demonstrate a very high and level values of specificity for \textit{H} and \textit{E} classes. For class \textit{C}, specificity is significantly improved via MCP2 and MCP5 respectively. Except for MCP1 in class \textit{C}, the values of specificity for all variations of our method and particularly in every class of the secondary structure is remarkably higher compared to the values of other metrics (e.g. MCC). Since the rate of true-negative compared to false-positive is large, our proposed method reaches a higher effectiveness.\\
As the MCC metric concurrently considers all the elements of the confusion matrix (i.e. True-Positive and etc), it can reflect reliable evaluation outcomes. The MCC is further useful for the class imbalance issue where the size of the classes - like in our dataset - is different. Based on the experiment, Figure \ref{fig:allMeasAlllClass}(d) shows high values of MCC for class \textit{H} and even values for \textit{E} and \textit{C} classes.\\
With regard to the classification task, the leveler and simultaneously higher values the evaluation metrics gain, the more effective the classifier will be. Hence, MCP5 and then MCP2 achieve the best performance. Moreover, the overall effectiveness of the advanced versions of our proposed approach (i.e. MCP2 and MCP5) are significantly promoted compared to the primary MCP1 model.
\vspace{-1mm}
\begin{table}[H]
	\footnotesize
	\centering
	\caption{Effectiveness on class H}
	\vspace{-2mm}
	\label{table:mcp H com}
	\begin{tabular}{p{5em}cccc}
		\toprule
		\textbf{Method} & \multicolumn{1}{p{6.8em}}{\textbf{Precision }} & \multicolumn{1}{p{4.5em}}{\textbf{Recall}} & \multicolumn{1}{p{6.355em}}{\textbf{Specificity}} & \multicolumn{1}{p{4.215em}}{\textbf{MCC}} \\
		\midrule
		MCP1 & \textbf{89.13} & \underline{61.02} & \textbf{97.74} & \underline{0.68 }\\[4pt]
		MCP2 & 85.61 & 72.92 & 96.28 & 0.73 \\[4pt]
		MCP5 & \underline{82.25} & \textbf{82.54} & \underline{94.59} & \textbf{0.77} \\[4pt]
		\bottomrule
	\end{tabular}%
	\label{tab:addlabel}%
\end{table}%
\vspace{-5mm}
\begin{table}[H]
	\footnotesize
	\centering
	\caption{Effectiveness on class E}
	\label{table:mcp E com}
	\begin{tabular}{p{5em}cccc}
		\toprule
		\textbf{Method} & \multicolumn{1}{p{6.8em}}{\textbf{Precision }} & \multicolumn{1}{p{4.5em}}{\textbf{Recall}} & \multicolumn{1}{p{6.355em}}{\textbf{Specificity}} & \multicolumn{1}{p{4.215em}}{\textbf{MCC}} \\
		\midrule
		MCP1 & \textbf{96.27} & \underline{81.87} & \textbf{98.53} & \underline{0.84} \\[4pt]
		MCP2 & 92.12 & 89.76 & 96.43 & 0.87 \\[4pt]
		MCP5 & \underline{90.91} & \textbf{94.22} & \underline{95.63} & \textbf{89} \\[4pt]
		\bottomrule
	\end{tabular}%
	\label{tab:addlabel}%
\end{table}%
\vspace{-5mm}
\begin{table}[H]
	\footnotesize
	\centering	
	\caption{Effectiveness on class C}
	\label{table:mcp C com}
	\begin{tabular}{p{5em}cccc}
		\toprule
		\textbf{Method} & \multicolumn{1}{p{6.8em}}{\textbf{Precision }} & \multicolumn{1}{p{4.5em}}{\textbf{Recall}} & \multicolumn{1}{p{6.355em}}{\textbf{Specificity}} & \multicolumn{1}{p{4.215em}}{\textbf{MCC}} \\
		\midrule
		MCP1 & \underline{75.18} & \textbf{95.38} & \underline{74.23} & \underline{0.7} \\[4pt]
		MCP2 & 81.12 & 88.81 & 83.09 & 0.72 \\[4pt]
		MCP5 & \textbf{87.05} & \underline{84.65} & \textbf{89.69} & \textbf{0.75} \\[4pt]
		\bottomrule
	\end{tabular}%
	\label{tab:addlabel}%
\end{table}
\vspace{-3mm}
Tables \ref{table:mcp H com}, \ref{table:mcp E com}, and \ref{table:mcp C com} compare MCP2 and MCP5 versus the initially devised classifier (MCP1). The results include the evaluation metrics of precision, recall, specificity, and MMC for the \textit{H}, \textit{E}, and \textit{C} classes. The maximum and minimum values are respectively highlighted with the bold and underlined formats.
For the class \textit{H} (Table \ref{table:mcp H com}) MCP1 owns the lowest values of recall and MCC and concurrently the highest values of precision and specificity. In an opposite manner, MCP5 has the lowest values of precision and specificity and the highest values of recall and MCC. The same scenario holds for class \textit{E} (Table \ref{table:mcp E com}). The behavior of the models is quite different towards class \textit{C} (Table \ref{table:mcp C com}) for which, MCP1 exhibits the lowest values of the precision, specificity, and MCC. Nevertheless, MCP5 acquires the highest values for these metrics. In total, MCP2 achieves moderate values for all the metrics in three classes. However, since MCP5 gains the highest MCC and $Q_3$ values for all three classes, it clearly outperforms other variations of our proposed framework.\\
Genuinely, there are 8 categories of protein secondary structures. The number of categories can be reduced to three by applying various reduction rules (e.g. DSSP \cite{Alirezaee2012}) on similar properties. According to Tables \ref{table:mcp H com}, \ref{table:mcp E com}, and \ref{table:mcp C com}, the measures exhibit higher values for class \textit{H} compared to other classes. The reason is that the diversity of the protein categories is less in class \textit{H} and it sufficiently includes one-third of the dataset samples.\\
\vspace{-9mm}
\subsection{Comparison and discussion}
\label{sec:evaluation}
\vspace{-4mm}
In this section, we compare the effectiveness of the variations of our framework in the prediction of protein secondary structure versus other baselines in the literature. The rival methods in this experiment are elucidated as follows:
\vspace{-2mm}
\begin{itemize}
	\item \textit{Muti-Component predictor with Aggregation 1}(MCP1): As explained in Section \ref{subsec:method}, MCP1 utilizes the Aggregation Rule 1 in the framework. Four other versions of our proposed solution are also devised through employing other aggregation rules. For instance, MCP2 utilizes the aggregation rule 2.	
	\item  \textit{Support Vector Regression Using the Non-Dominated Sorting Genetic Algorithm \Romannum{2} and Dynamic Weighted Kernel Fusion} (SVR-NSGA\Romannum{2}): This model proposed by Zangooei et al. \cite{Zangooei2012} takes sequence profiles (numerical vectors) as input. This baseline on the one hand, employs the support vector regression for the classification task, and on the other hand utilizes NSGA\Romannum{2}\ to map the real values to integers and subsequently optimize the kernel parameters. The model also applies the weighted fusion of three kernel functions.
	\item  \textit{Support Vector Machine using Parallel Hierarchical Grid Search and Dynamic Weighted Kernel Fusion} (SVM-PHGS): The model proposed in \cite{Zangooei2012a} takes the sequence profiles (numerical vectors of evolutionary information) as the training input. The baseline method utilizes SVM with a compound kernel function and PHGS to optimize the kernel parameters.
	\item  \textit{Ensemble Method Using Neural Networks and Support Vector Machines}(EM): \cite{Bouziane2015} is a method that combines the vote of multiple individual learners comprising a multi-layer perceptron (MLP), an RBF neural network, and four SVM classifiers. The method applies numerous combination rules to generate the final prediction output. Bouziane et al. \cite{Bouziane2015} also investigated the effect of two input data types including Position-Specific Scoring Matrix (PSSM) \cite{Ng2001} and a coding scheme which is used to represent the amino-acids.	
	\item  \textit{Support Vector Machine Using Hybrid Coding for Protein} (SVM-HC): 
	The method proposed by Li et al. \cite{Li2017} initially employs the geometry-based similarities and where the similarity comparison is not applicable, the SVM module will be used to leverage the final protein structure. Moreover, SVM-HC extracts a 6-bit code from the physiochemical properties of both amino-acids and the tendency factors.	
\end{itemize} 
In this experiment, we use the RS126, CASP11 and CASP12 datasets to compare different versions of our approach with other rivals. The competitors have also similar learning modules. For instance, all the frameworks of SVR-NSGA\Romannum{2}, SVM-PHGS, and EM utilize the SVM to accomplish the learning task. Like our proposed framework, the EM approach accommodates various combination rules. Moreover, the SVM-HC as an SVM based approach utilizes the similarity metrics for the first phase of the predictions. The abbreviation for each version of our proposed approach is numbered based on the aggregation rule - e.g. MCP1 represents the Multi-component predictor with aggregation Rule 1.
\begin{table}[H]
	\footnotesize
	\centering
	\caption{Effectiveness of structure prediction - Various versions of the baselines}
	\vspace{-4mm}
	\label{tabe:baseline}
	\label{table:baseline}
	\begin{tabular}{p{12em}:c:c:c:c:c:c:c}
		\toprule
		\textbf{Method Name} & \multicolumn{1}{p{2.5em}}{\textbf{$Q_3$}} & \multicolumn{1}{p{2.5em}}{\textbf{$Q_H$}} & \multicolumn{1}{p{2.5em}}{\textbf{$Q_E$}} & \multicolumn{1}{p{2.5em}}{\textbf{$Q_C$}} & \multicolumn{1}{p{2.5em}}{\textbf{$MCC_H$}} & \multicolumn{1}{p{2.5em}}{\textbf{$MCC_E$}} & \multicolumn{1}{p{2.5em}}{\textbf{$MCC_C$}} \\
		\midrule
		EM\_EXPOP(SC) \cite{Bouziane2015} & 65.2  & 61.54 & 40.76 & 78.8  & 0.473 & 0.387 & 0.438 \\
		EM\_LOGOP(SC) \cite{Bouziane2015} & 65.19 & 61.59 & 40.78 & 78.73 & 0.474 & 0.387 & 0.438 \\
		EM\_LINOP(SC) \cite{Bouziane2015} & 65.19 & 61.59 & 40.78 & 78.73 & 0.474 & 0.386 & 0.438 \\
		EM\_EXPOP(PSSM) \cite{Bouziane2015} & 78.14 & 77.18 & 65.34 & 84.86 & 0.72  & 0.624 & 0.615 \\
		EM\_LOGOP(PSSM) \cite{Bouziane2015} & 78.12 & 77.2  & 65.28 & 84.83 & 0.72  & 0.624 & 0.615 \\
		EM\_LINOP(PSSM) \cite{Bouziane2015} & 78.14 & 77.18 & 65.34 & 84.86 & 0.72  & 0.624 & 0.615 \\
		\midrule
		SVM-PHGS(DWKF) \cite{Zangooei2012a} & 84.6  & 91.2  & 72.1  & 84.3  & NA & NA & NA \\
		SVR-NSGAII(DWKF) \cite{Zangooei2012} & 85.75 & 92.47 & 78.41 & 85.11 & NA & NA & NA \\
		\midrule
		SVM\_HC \cite{Li2017} & 82.5  & 82.1  & 65.09 & 89.07 & 0.779  & 0.761  & 0.748 \\
		\midrule
		MCP1  & 83.09 & 81.87 & 61.02 & \textbf{95.4} & 0.84  & 0.68  & 0.7 \\
		MCP2  & 85.41 & 89.76 & 72.92 & 88.81 & 0.87  & 0.73  & 0.72 \\
		MCP5  & \textbf{87.3} & \textbf{94.2} & \textbf{82.5} & 84.65 & \textbf{0.89} & \textbf{0.77} & \textbf{0.75} \\
		\bottomrule
	\end{tabular}%
\end{table}
\vspace{-3mm}
Table \ref{table:baseline} offers a comparison between the introduced baseline methods. Our approach owns the highest values for all evaluation metrics. The two SVM-based methods named as SVR-NSGA\Romannum{2} and SVM-PHGS employ support vector machine besides a weighted fusion of three kernels. Also, they optimize the kernel parameters through the grid search and genetic algorithms. However, compared to the aforementioned SVM-based models, even without any optimization, our proposed multi-component framework achieves a better performance. The reason lies in the fact that the multi-component architecture can better address the sequence-structure complexity. The table lists various versions of the EM method including \textit{Sequence Coding} (postfixed with SC e.g. EM\_EXPOP) and Position-Specific Scoring Matrix (PSSM). Furthermore, despite the fact that the EM method as an ensemble SVM-based machine takes advantage of several SVM modules and introduces a variety of combiners, our approach still can notably outperform various variations of this baseline. The EM baseline suffers from two possible deficiencies. First, the insufficient diversity among SVM components which is a necessary property of an ensemble system. Second, the use of the encoding schemes which can result in information loss. Additionally, our approach excels the SVM-HC baseline. Similar to our Fuzzy classification component (FKNN), SVM-HC initially employs a similarity metric for structure prediction and subsequently applies an SVM classifier on the residues that are not classified in the initial step. What empowers our method versus SVM-HC is that our approach further employs the effective compound dissimilarity measure ($\widetilde{d}$) to directly process the protein sequences and further uses an efficient fuzzy aggregation component. Tables \ref{table:literatureQ_com} and \ref{table:literatureMCC} compare the effectiveness of our approach versus other baselines as discussed in the literature (Section \ref{sec:relWork}). Table \ref{table:literatureQ_com} demonstrates how our approach outperforms various learners including the ensemble machines, neural networks, SVMs, and decision trees. Since our method uses the fuzzy aggregation process besides a multi-component designation, it outperforms other baselines. Table \ref{table:distance-based methods} further compares the effectiveness of other distance-based classifiers against our FKNN-based extensions using the dissimilarity parameters (i.e. LZ scores, n-gram scores, and $\rho_{d}$). The results in Table \ref{table:distance-based methods} prove that taking the string protein sequences for the input - as utilized in our FKNN-based components - is more beneficial compared to consuming of the numerical encoding schemes as used in rival baselines. Furthermore, the results clarify that our dissimilarity measure can better improve prediction outcomes compared to the distance metrics which are used in the rival methods.
\vspace{-4mm}
\begin{table}[H]
	\footnotesize
	\centering
	\begin{tabular}{p{20.855em}:c:c:c:c}
		\toprule
		\textbf{Method's Name} & \multicolumn{1}{p{2.5em}}{\textbf{$Q_3$}} & \multicolumn{1}{p{2.5em}}{\textbf{$Q_H$}} & \multicolumn{1}{p{2.5em}}{\textbf{$Q_E$}} & \multicolumn{1}{p{2.5em}}{\textbf{$Q_C$}} \\
		\midrule
		Ensemble NN (Imbalanced Training Set) \cite{Alirezaee2012} & 73.33 & 68.07 & 52.78 & 73.02 \\
		Ensemble NN (Over-Sampling) \cite{Alirezaee2012} & 73.75 & 71.72 & 68.9  & 77.44 \\
		Ensemble NN (Under-Sampling) \cite{Alirezaee2012} & 73.02 & 69.76 & 77.48 & 73.2 \\
		Ensemble NN (Under and Over-Sampling)\cite{Alirezaee2012}  & 73.73 & 71.05 & 69.38 & 77.38 \\
		Ensemble NN (Tree-based) \cite{Alirezaee2012} & 73.51 & 75.78 & 67.85 & 74.84 \\
		NN with SMV voting scheme \cite{Alirezaee2012} & 74.66 & 74.85 & 72.78 & 75.32 \\
		NN with GWMV voting scheme \cite{Alirezaee2012} & 74.9  & 72.61 & 70.25 & 78.56 \\
		NN with WMV voting scheme \cite{Alirezaee2012} & 74.64 & 72.43 & 69.76 & 78.43 \\
		\midrule
		PLM-PBC-HPP \cite{Wang2008} & 69    & 67.1  & 62.7  & 73.4 \\
		ELM-HPP01 \cite{Wang2008} & 67.9  & 62.5  & 61.9  & 71.6 \\
		Single-Stage One-against-all \cite{Nguyen2003} & 69.7  & 54.1  & 79.3  & 59 \\
		Single-Stage One-against-one \cite{Nguyen2003} & 67.6  & 54.5  & 79.8  & 58.3 \\
		Single-Stage DAG \cite{Nguyen2003} & 67.5  & 54.2  & 80    & 58.3 \\
		Single-Stage Crammer and Singer \cite{Nguyen2003} & 70.2  & 55.8  & 78    & 56.5 \\
		Two-Stage One-against-all \cite{Nguyen2003} & 66.5  & 61.2  & 78.5  & 63.9 \\
		Two-Stage One-against-one \cite{Nguyen2003} & 66.5  & 57.5  & 81.2  & 65.4 \\
		Two-Stage DAG \cite{Nguyen2003} & 66.8  & 57.4  & 80.9  & 65.5 \\
		Single-Stage Vapnik and Weston \cite{Nguyen2003} & 70.4  & 55.7  & 78.2  & 57.1 \\
		Two-Stage Vapnik and Weston \cite{Nguyen2003} & 66.1  & 57.8  & 81.9  & 67 \\
		Multi-SVM Ensemble \cite{Nguyen2003} & 74.98 & 75.37 & 66.43 & 79.26 \\
		Two-Stage Crammer and Singer \cite{Nguyen2003} & 66.8  & 57.9  & 81    & 66.8 \\
		\midrule
		DT \cite{He2006} & NA & 70.4  & 78.4  & 67.1 \\
		SVM-DT \cite{He2006} & NA & 72.8  & 79.6  & 69.3 \\
		\midrule
		MCP1  & 83.09 & 81.87 & 61.02 & \textbf{95.38} \\
		MCP2  & 85.41 & 89.76 & 72.92 & 88.81 \\
		MCP5  & \textbf{87.29} & \textbf{94.22} & \textbf{82.54} & 84.65 \\
		\bottomrule
	\end{tabular}
\vspace{-3mm}
\caption{The accuracy of our approach versus other competitors}
\label{table:literatureQ_com}	
\end{table}
\vspace{-8mm}
\begin{table}[H]
	\footnotesize
	\centering
	\begin{tabular}{p{24.855em}:c:c:c}
		\toprule
		\textbf{Method's Name} & \multicolumn{1}{p{2.5em}}{\textbf{$MCC_H$}} & \multicolumn{1}{p{2.5em}}{\textbf{$MCC_E$}} & \multicolumn{1}{p{2.6em}}{\textbf{$MCC_C$}} \\
		\midrule
		Ensemble NN (Imbalanced Training Set) \cite{Alirezaee2012} & 0.65  & 0.54  & 0.51 \\
		Ensemble NN (Over-Sampling) \cite{Alirezaee2012} & 0.66  & 0.57  & 0.51 \\
		Ensemble NN (Under-Sampling) \cite{Alirezaee2012} & 0.64  & 0.56  & 0.49 \\
		Ensemble NN (Under-Sampling and Over-Sampling) \cite{Alirezaee2012} & 0.65  & 0.56  & 0.51 \\
		Ensemble NN (Tree-based) \cite{Alirezaee2012} & 0.64  & 0.55  & 0.52 \\
		NN with SMV voting scheme \cite{Alirezaee2012} & 0.67  & 0.58  & 0.53 \\
		NN with GWMV voting scheme\cite{Alirezaee2012} & 0.67  & 0.58  & 0.53 \\
		NN with WMV voting scheme \cite{Alirezaee2012} & 0.67  & 0.58  & 0.53 \\
		\midrule
		MCP1  & 0.84  & 0.68  & 0.7 \\
		MCP2  & 0.87  & 0.73  & 0.72 \\
		MCP5  & \textbf{0.89} & \textbf{0.77} & \textbf{0.75} \\
		\bottomrule
	\end{tabular}
	\vspace{-3mm}
	\caption{The comparison of the MCC metric using the RS126 dataset}
	\label{table:literatureMCC}
\end{table}%

\begin{figure}[H]
	\footnotesize
	\centering
	\vspace{-2mm}
	\caption{Comparison of the accuracy and MCC between the best baselines}
	\includegraphics[trim=0cm 0cm 0cm 0cm,clip=true,scale=0.48]{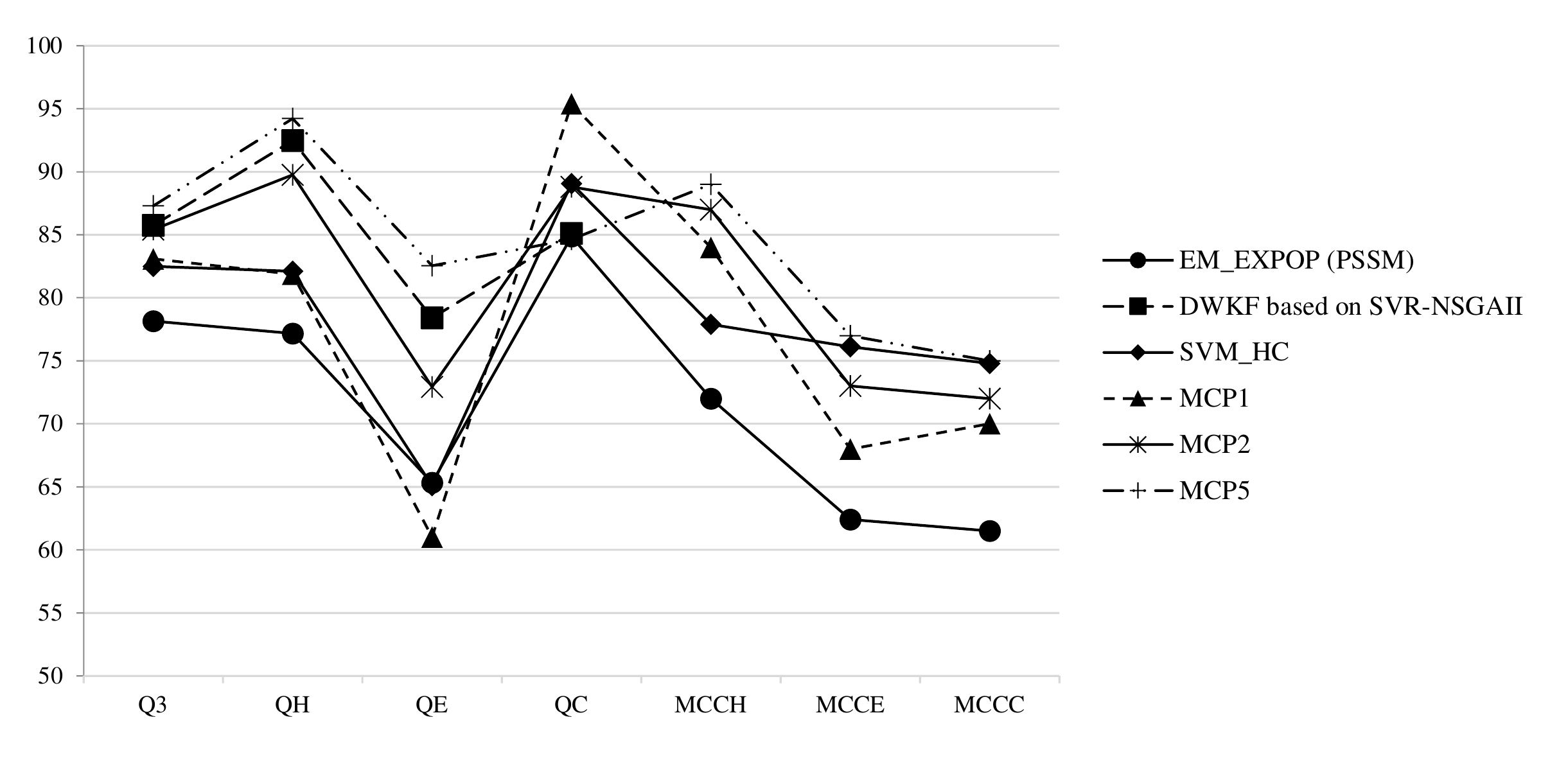}
	\label{fig:baseline trend}
	\end{figure}

\begin{table}[H]
	\footnotesize
	\centering
	\vspace{-2mm}
	\caption{The $Q_3$ metric for the distance-based methods}
	\label{table:distance-based methods}
	\begin{tabular}{p{13.215em}:c}
		\toprule
		\textbf{Method's Name} & \multicolumn{1}{p{4.215em}}{\textbf{$Q_3$}} \\
		\midrule
		KNN \cite{Ghosh2008} & 49.85 \\
		Fuzzy KNN \cite{Ghosh2008} & 53.08 \\
		Minimum Distance \cite{Ghosh2008} & 59.22 \\
		\midrule
		FKNN + LZ scores & 76.38 \\
		FKNN + LZ + $\rho_d$ & 80.42 \\
		FKNN + LZ + $\rho_d$ + n-gram & \textbf{81.96} \\
		\bottomrule
	\end{tabular}%
\end{table}
\vspace{-3mm}
In terms of protein secondary structure prediction, Figure \ref{fig:baseline trend} shows the performance comparison between the best version of each of the baselines versus three variations of our proposed approach (MCP1, MCP2, and MCP5).
\vspace{-4mm}
	\vspace{-4mm}
As shown in the figure, EM\_EXPOP(PSSM) has an overall poor performance across all classes. The MCP5 model gains the highest and the evenest values for all measures on two major classes of \textit{E} and \textit{H}. While MCP5 can effectively predict the protein structure in class \textit{E}, almost all other competitors obtain a very low accuracy in this class.
\vspace{-8mm}
\subsection{Conclusion}
\label{sec:conclusion}
\vspace{-4mm}
Given the input protein sequences, in this paper, we propose a unified multi-component framework to predict protein secondary structure effectively. More specifically, we divide the prediction task into four subtasks: Pre-processing, classification, filtration, and aggregation. We utilize a sliding-window approach for the pre-processing task to make the sequences equal in length and improve the results by including the neighboring residues. We facilitate the classification module through employing two classifiers of Edit-SVM and $\widetilde{d}$-FKNN. The fuzzy component uses three dissimilarity measures of LZ score, n-gram score, and the dissimilarity rate which are fused into a unified dissimilarity metric. Subsequently, the output of classification module will be filtered biologically to eliminate meaningless structures. The refined output will then be fed into the aggregation pool which comprises multiple rules to better infer the outcome secondary structure. Additionally, our method is capable to directly process string protein sequences and avoid feature extraction. Feature extraction can cause data loss and increase the input data dimensions. Additionally, we consume the proteins input sequences that can globally exploit the latent natural language and better explain the hidden relationship between the input sequence and the extracted output structures. Processing textual contents is also beneficial to overcome the negative influence of the high dimensional numerical feature vectors. The experimental results of our approach and the numerical feature-based methods demonstrate the strength of the string dissimilarity measures. Generally, our approach performs prediction effectively and compared to other rivals enhances the overall accuracy of the prediction process.\\
\vspace{-8mm}
\subsection{Future work}
\label{sec:Future work}
\vspace{-4mm}
Our multi-component approach is flexible and its modules can be replaced and modified. As the initial change, one might want to add more classifiers and expand the ensemble size with different base learners. Fuzzification of Edit-SVM module can enhance the aggregation process and consequently enrich the final prediction results. Moreover, a weighted dissimilarity measure in Fuzzy KNN can better validate the role for each of the dissimilarity components and result in a higher accuracy. Additionally, a parameter optimization approach can adjust the parameters in the prediction process. Hence, an extensive range of modifications can be applied to our unified multi-component framework. We leave  such modifications as future work. Additionally, a user-friendly and publicly accessible protein secondary structure prediction web-servers facilitates the development of practically more useful prediction methods and computational tools. Actually, many practically useful web-servers have significantly contributed to the advancement of medical science and chemistry. Consequently, we shall make efforts in our future work to provide a web-server based on the prediction method presented in this paper\\
\vspace{-12mm}
\section*{Author Biographies}
\vspace{-3mm}
\leavevmode
\vbox{%
\begin{wrapfigure}{l}{50pt}
{\vspace*{-10pt}\fbox{\includegraphics[scale=0.35]{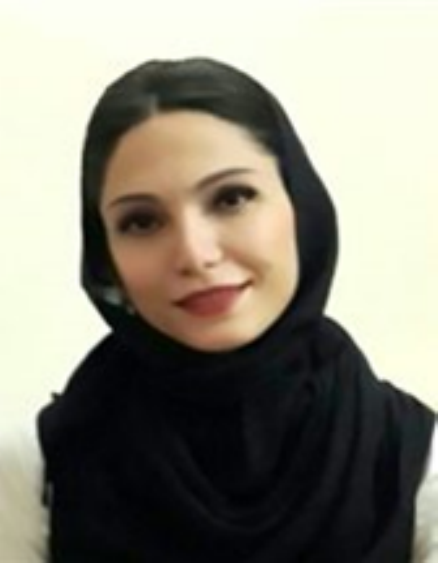}}\vspace*{100pt}}%
\end{wrapfigure}
\noindent\small 
{\bf Leila Khalatbari} is currently a Ph.D. candidate in Artificial Intelligence at Sharif University of Technology, Tehran, Iran. Since 2014, she has been a researcher in the computational cognitive models laboratory at Iran University of Science and Technology. She received her B.S. degree in computer software engineering from Qazvin. She recently completed her M.S. degree in Artificial Intelligence. Her research interests include machine learning, data analysis, data mining, and bioinformatics.\vadjust{\vspace{20pt}}}

\vbox{%
\begin{wrapfigure}{l}{50pt}
{\vspace*{-10pt}\fbox{\includegraphics[scale=0.35]{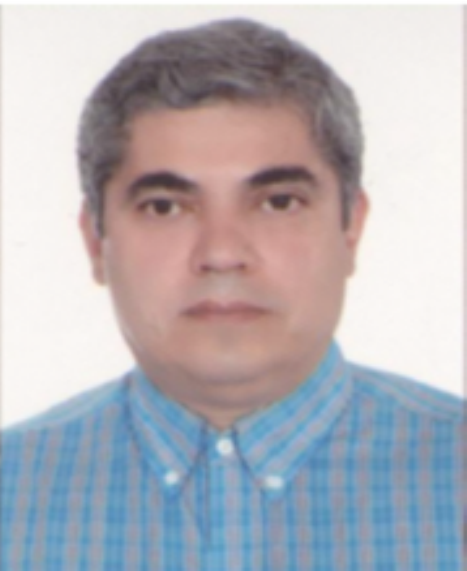}}\vspace*{100pt}}%
\end{wrapfigure}
\noindent\small {\bf Mohammad Reza Kangavari} received the B.S. degree in mathematics and computer science from the Sharif University of Technology in 1982, the M.S. degree in computer science from Salford University in 1989, and the Ph.D. degree in computer science from the University of Manchester in 1994. He is currently a lecturer in the Computer Engineering Department, Iran University of Science and Technology, Tehran, Iran. His research interests include Intelligent Systems, Human Computer-Interaction, Cognitive Computing, Data Science, Machine Learning, and Wireless Sensor Networks.
\vadjust{\vspace{10pt}}}
\vbox{%

\begin{wrapfigure}{l}{50pt}
{\vspace*{-10pt}\fbox{\includegraphics[scale=0.12]{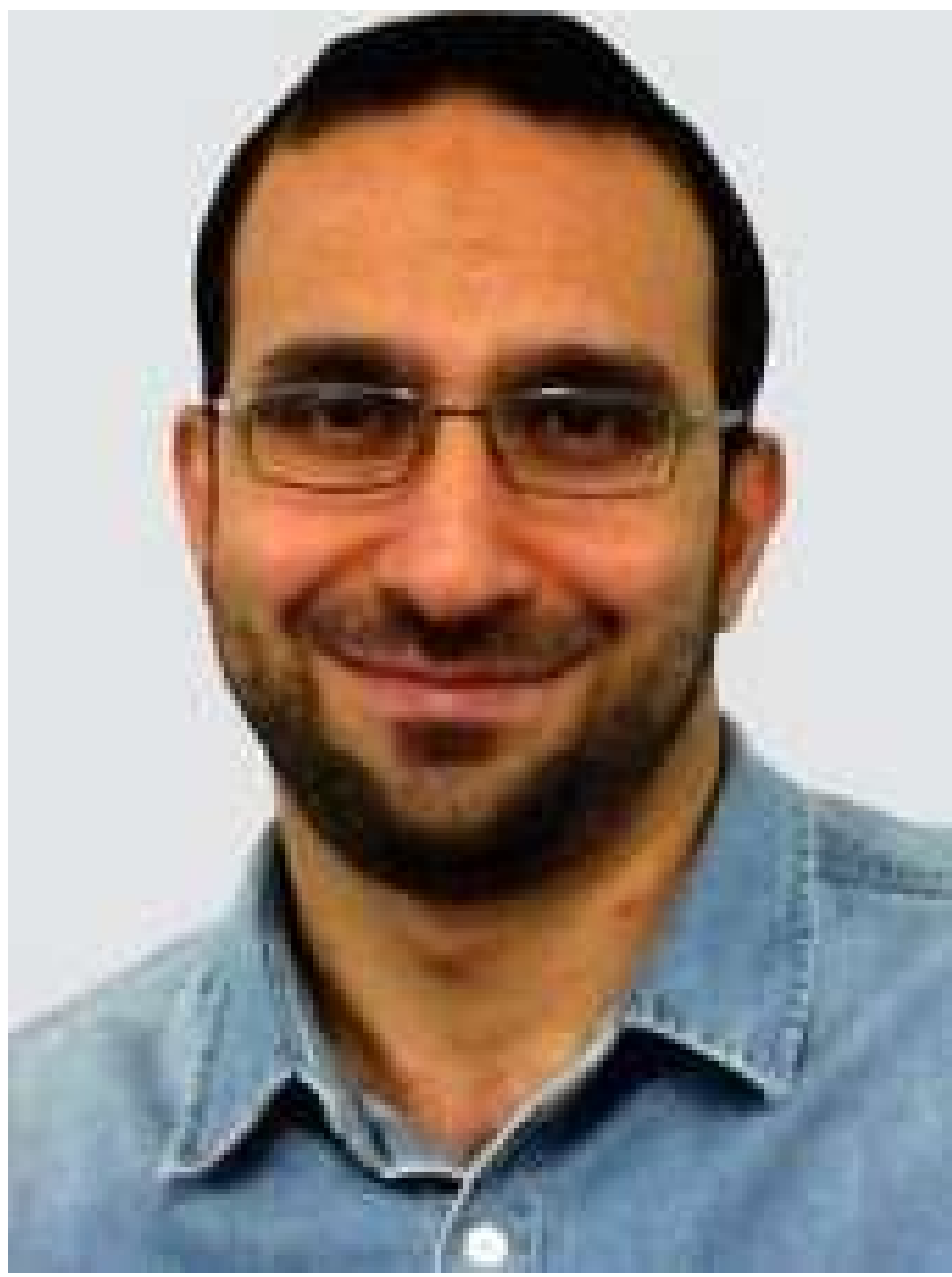}}\vspace*{100pt}}%
\end{wrapfigure}
\noindent\small 
{\bf Saeid Hosseini} completed the Ph.D. degree in Computer Science at the University of Queensland, Australia. He obtained his M.Sc. from the Queensland University Of Technology in 2012. He received the Australian Postgraduate Award in 2015. He is currently a post-doc researcher. His research interests mainly focus on, diffusion networks, graph mining, crowdsourcing, recommendation systems, spatiotemporal databases, and social network analytics. Dr. Hosseini is an invited reviewer in reputed proceedings including ICDM and TKDE. He has also been a program committee member in CSS (2017) and DASFAA (2017 and 2018).\vadjust{\vspace{10pt}}
}
\vbox{%
\begin{wrapfigure}{l}{50pt}
{\vspace*{-10pt}\fbox{\includegraphics[scale=0.28]{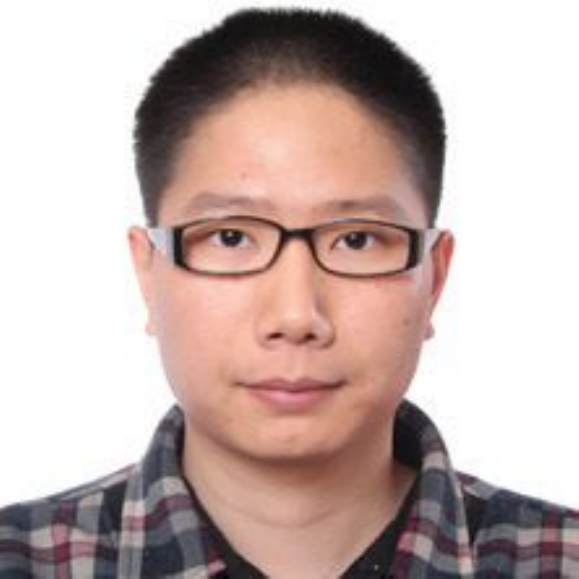}}\vspace*{100pt}}%
\end{wrapfigure}
\noindent\small 
{\bf Hongzhi Yin} works as a Lecturer and an ARC DECRA Fellow with The University of Queensland, Australia. He received his doctoral degree from Peking University in July 2014 under the supervision of Prof. Bin Cui. His research interests include user behavior modeling, user profiling,  recommender system, especially spatial-temporal recommendation, user linkage across social networks, network embedding, knowledge graph mining, and construction, topic discovery and event detection, deep learning.  He has been serving as conference organizers, conference PC member for PVLDB, SIGIR, ICDE, IJCAI, ICDM, CIKM, DASFAA, ASONAM, MDM, WISE, PAKDD and reviewer of more than 10 reputed journals such as VLDB Journal, TKDE, TOIS, TKDD, and etc.\vadjust{\vspace{10pt}}}

\vbox{%
	\begin{wrapfigure}{l}{50pt}
		{\vspace*{-10pt}\fbox{\includegraphics[scale=0.40]{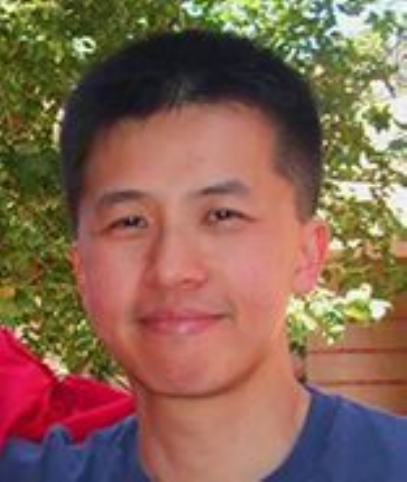}}\vspace*{100pt}}%
	\end{wrapfigure}
	\noindent\small 
	{\bf Ngai-Man Cheung} received the Ph.D. degree in electrical engineering from the University of Southern California, Los Angeles, CA, in 2008. He is currently an Assistant Professor with the Singapore University of Technology and Design (SUTD). From 2009-2011, he was a postdoctoral researcher with the Image, Video and Multimedia Systems group at Stanford University, Stanford, CA.  He has also held research positions with Texas Instruments Research Center Japan, Nokia Research Center, IBM T. J. Watson Research Center, HP Labs Japan, Hong Kong University of Science and Technology (HKUST), and Mitsubishi Electric Research Labs (MERL). His work has resulted in 11 U.S. patents granted with several pending. His research interests include signal, image, and video processing, computer vision and machine learning.}
\label{lastpage}

\vspace{-5mm}
\bibliographystyle{abbrv}
\bibliography{location_modelling}

\begin{thebibliography}{10}

\bibitem{abachi2018statistical}
H.~M. Abachi, S.~Hosseini, M.~A. Maskouni, M.~Kangavari, and N.-M. Cheung.
\newblock Statistical discretization of continuous attributes using
  kolmogorov-smirnov test.
\newblock pages 309--315, 2018.

\bibitem{Alirezaee2012}
M.~Alirezaee, A.~Dehzangi, and E.~Mansoori.
\newblock Ensemble of neural networks to solve class imbalance problem of
  protein secondary structure prediction.
\newblock {\em International Journal of Artificial Intelligence \&
  Applications}, 3(6):9, 2012.

\bibitem{Aygun2010}
E.~Aygun, B.~J. Oommen, and Z.~Cataltepe.
\newblock Peptide classification using optimal and information theoretic
  syntactic modeling.
\newblock {\em Pattern Recognition}, 43(11):3891--3899, 2010.

\bibitem{Babaei2010}
S.~Babaei, A.~Geranmayeh, and S.~A. Seyyedsalehi.
\newblock Protein secondary structure prediction using modular reciprocal
  bidirectional recurrent neural networks.
\newblock {\em Computer methods and programs in biomedicine}, 100(3):237--247,
  2010.

\bibitem{Babaei2012}
S.~Babaei, A.~Geranmayeh, and S.~A. Seyyedsalehi.
\newblock Towards designing modular recurrent neural networks in learning
  protein secondary structures.
\newblock {\em Expert Systems with Applications}, 39(6):6263--6274, 2012.

\bibitem{Bondugula2005}
R.~Bondugula, O.~Duzlevski, and D.~Xu.
\newblock Profiles and fuzzy k-nearest neighbor algorithm for protein secondary
  structure prediction.
\newblock In {\em Proceedings of the 3rd Asia-Pacific Bioinformatics
  Conference}, pages 85--94. World Scientific, 2005.

\bibitem{Bouziane2015}
H.~Bouziane, B.~Messabih, and A.~Chouarfia.
\newblock Effect of simple ensemble methods on protein secondary structure
  prediction.
\newblock {\em Soft Computing}, 19(6):1663--1678, 2015.

\bibitem{Bystroff2008}
C.~Bystroff and A.~Krogh.
\newblock Hidden markov models for prediction of protein features.
\newblock In {\em Protein Structure Prediction}, pages 173--198. Springer,
  2008.

\bibitem{Chen2007}
C.~Chen, Y.~Tian, X.~Zou, P.~Cai, and J.~Mo.
\newblock Prediction of protein secondary structure content using support
  vector machine.
\newblock {\em Talanta}, 71(5):2069--2073, 2007.

\bibitem{Chen2017}
Y.~Chen.
\newblock Long sequence feature extraction based on deep learning neural
  network for protein secondary structure prediction.
\newblock In {\em Information Technology and Mechatronics Engineering
  Conference (ITOEC), 2017 IEEE 3rd}, pages 843--847. IEEE, 2017.

\bibitem{Chou1974}
P.~Y. Chou and G.~D. Fasman.
\newblock Prediction of protein conformation.
\newblock {\em Biochemistry}, 13(2):222--245, 1974.

\bibitem{Dinubhai2014}
P.~M. Dinubhai and H.~B. Shah.
\newblock Protein secondary structure prediction using neural network: a
  comparative study.
\newblock {\em Int J Enhanc Res Manag Comput Appl}, 3(4):18--23, 2014.

\bibitem{Fang2018}
C.~Fang, Y.~Shang, and D.~Xu.
\newblock Mufold-ss, new deep inception-inside-inception networks for protein
  secondary structure prediction.
\newblock {\em Proteins: Structure, Function, and Bioinformatics}, 2018.

\bibitem{Garnier1996}
J.~Garnier, J.-F. Gibrat, and B.~Robson.
\newblock Gor method for predicting protein secondary structure from amino acid
  sequence.
\newblock In {\em Methods in enzymology}, volume 266, pages 540--553. Elsevier,
  1996.

\bibitem{Garnier1978}
J.~Garnier, D.~J. Osguthorpe, and B.~Robson.
\newblock Analysis of the accuracy and implications of simple methods for
  predicting the secondary structure of globular proteins.
\newblock {\em Journal of molecular biology}, 120(1):97--120, 1978.

\bibitem{Ghosh2008}
A.~Ghosh and B.~Parai.
\newblock Protein secondary structure prediction using distance based
  classifiers.
\newblock {\em International journal of approximate reasoning}, 47(1):37--44,
  2008.

\bibitem{Han2011}
J.~Han, J.~Pei, and M.~Kamber.
\newblock {\em Data mining: concepts and techniques}.
\newblock Elsevier, 2011.

\bibitem{He2006}
J.~He, H.-J. Hu, R.~Harrison, P.~C. Tai, and Y.~Pan.
\newblock Rule generation for protein secondary structure prediction with
  support vector machines and decision tree.
\newblock {\em IEEE Transactions on nanobioscience}, 5(1):46--53, 2006.

\bibitem{hosseini2014location}
S.~Hosseini, S.~Unankard, X.~Zhou, and S.~Sadiq.
\newblock Location oriented phrase detection in microblogs.
\newblock pages 495--509, 2014.

\bibitem{hosseini2018exploiting}
S.~Hosseini, H.~Yin, N.-M. Cheung, K.~P. Leng, Y.~Elovici, and X.~Zhou.
\newblock Exploiting reshaping subgraphs from bilateral propagation graphs.
\newblock pages 342--351, 2018.

\bibitem{Hosseini}
S.~Hosseini, H.~Yin, M.~Zhang, Y.~Elovici, and X.~Zhou.
\newblock Mining subgraphs from propagation networks through temporal dynamic
  analysis.

\bibitem{Hosseini2017a}
S.~Hosseini, H.~Yin, M.~Zhang, X.~Zhou, and S.~Sadiq.
\newblock Jointly modeling heterogeneous temporal properties in location
  recommendation.
\newblock In {\em International Conference on Database Systems for Advanced
  Applications}, pages 490--506. Springer, 2017.

\bibitem{Hosseini2017}
S.~Hosseini, H.~Yin, X.~Zhou, S.~Sadiq, M.~R. Kangavari, and N.-M. Cheung.
\newblock Leveraging multi-aspect time-related influence in location
  recommendation.
\newblock {\em World Wide Web}, pages 1--28, 2017.

\bibitem{Hua2012}
W.~Hua, D.~T. Huynh, S.~Hosseini, J.~Lu, and X.~Zhou.
\newblock Information extraction from microblogs: A survey.
\newblock {\em Int. J. Software and Informatics}, 6(4):495--522, 2012.

\bibitem{Johal2014}
A.~K. Johal and R.~Singh.
\newblock Protein secondary structure prediction using improved support vector
  machine and neural networks.
\newblock {\em International Journal of Engineering and Computer Science},
  3(1):3593--3597, 2014.

\bibitem{Keller1985}
J.~M. Keller, M.~R. Gray, and J.~A. Givens.
\newblock A fuzzy k-nearest neighbor algorithm.
\newblock {\em IEEE transactions on systems, man, and cybernetics},
  (4):580--585, 1985.

\bibitem{Krissinel2007}
E.~Krissinel.
\newblock On the relationship between sequence and structure similarities in
  proteomics.
\newblock {\em Bioinformatics}, 23(6):717--723, 2007.

\bibitem{Lee2009}
S.~Y. Lee, J.~Y. Lee, K.~S. Jung, and K.~H. Ryu.
\newblock A 9-state hidden markov model using protein secondary structure
  information for protein fold recognition.
\newblock {\em Computers in biology and medicine}, 39(6):527--534, 2009.

\bibitem{Lempel1976}
A.~Lempel and J.~Ziv.
\newblock On the complexity of finite sequences.
\newblock {\em IEEE Transactions on information theory}, 22(1):75--81, 1976.

\bibitem{Li2017}
Z.~Li, J.~Wang, S.~Zhang, Q.~Zhang, and W.~Wu.
\newblock A new hybrid coding for protein secondary structure prediction based
  on primary structure similarity.
\newblock {\em Gene}, 618:8--13, 2017.

\bibitem{Lin2010}
H.-N. Lin, T.-Y. Sung, S.-Y. Ho, and W.-L. Hsu.
\newblock Improving protein secondary structure prediction based on short
  subsequences with local structure similarity.
\newblock In {\em Bmc Genomics}, volume~11, page~S4. BioMed Central, 2010.

\bibitem{Lin2010a}
L.~Lin, S.~Yang, and R.~Zuo.
\newblock Protein secondary structure prediction based on multi-svm ensemble.
\newblock In {\em Intelligent Control and Information Processing (ICICIP), 2010
  International Conference on}, pages 356--358. IEEE, 2010.

\bibitem{Liu2010}
T.~Liu, X.~Zheng, and J.~Wang.
\newblock Prediction of protein structural class using a complexity-based
  distance measure.
\newblock {\em Amino acids}, 38(3):721--728, 2010.

\bibitem{Liu2017}
Y.~Liu, J.~Cheng, Y.~Ma, and Y.~Chen.
\newblock Protein secondary structure prediction based on two dimensional deep
  convolutional neural networks.
\newblock In {\em Computer and Communications (ICCC), 2017 3rd IEEE
  International Conference on}, pages 1995--1999. IEEE, 2017.

\bibitem{maskouni2018auto}
M.~A. Maskouni, S.~Hosseini, H.~M. Abachi, M.~Kangavari, and X.~Zhou.
\newblock Auto-ces: An automatic pruning method through clustering ensemble
  selection.
\newblock pages 275--287, 2018.

\bibitem{Masulli2009}
F.~Masulli and S.~Mitra.
\newblock Natural computing methods in bioinformatics: A survey.
\newblock {\em Information Fusion}, 10(3):211--216, 2009.

\bibitem{Mossos2014}
N.~Mossos, D.~F. Mejia-Carmona, and I.~Tischer.
\newblock Fs-tree: Sequential association rules and first applications to
  protein secondary structure analysis.
\newblock In {\em Advances in Computational Biology}, pages 189--198. Springer,
  2014.

\bibitem{Ng2001}
P.~C. Ng and S.~Henikoff.
\newblock Predicting deleterious amino acid substitutions.
\newblock {\em Genome research}, 11(5):863--874, 2001.

\bibitem{Nguyen2003}
M.~N. Nguyen and J.~C. Rajapakse.
\newblock Multi-class support vector machines for protein secondary structure
  prediction.
\newblock {\em Genome Informatics}, 14:218--227, 2003.

\bibitem{Paliwal2015}
K.~Paliwal, J.~Lyons, and R.~Heffernan.
\newblock A short review of deep learning neural networks in protein structure
  prediction problems.
\newblock {\em Advanced Techniques in Biology \& Medicine}, pages 1--2, 2015.

\bibitem{Patel2014}
M.~S. Patel and H.~S. Mazumdar.
\newblock Knowledge base and neural network approach for protein secondary
  structure prediction.
\newblock {\em Journal of theoretical biology}, 361:182--189, 2014.

\bibitem{Pollastri2007}
G.~Pollastri, A.~J.~M. Martin, C.~Mooney, and A.~Vullo.
\newblock Accurate prediction of protein secondary structure and solvent
  accessibility by consensus combiners of sequence and structure information.
\newblock {\em BMC bioinformatics}, 8(1):201, 2007.

\bibitem{Rost1993}
B.~Rost and C.~Sander.
\newblock Prediction of protein secondary structure at better than 70%
  accuracy.
\newblock {\em Journal of molecular biology}, 232(2):584--599, 1993.

\bibitem{Spencer2015}
M.~Spencer, J.~Eickholt, and J.~Cheng.
\newblock A deep learning network approach to ab initio protein secondary
  structure prediction.
\newblock {\em IEEE/ACM transactions on computational biology and
  bioinformatics}, 12(1):103--112, 2015.

\bibitem{Tan2006}
P.-N. Tan.
\newblock {\em Introduction to data mining}.
\newblock Pearson Education India, 2006.

\bibitem{Tan2015}
Y.~T. Tan and B.~A. Rosdi.
\newblock Fpga-based hardware accelerator for the prediction of protein
  secondary class via fuzzy k-nearest neighbors with lempel-ziv complexity
  based distance measure, 2015.

\bibitem{Theodoridis2003}
S.~Theodoridis and K.~Koutroumbas.
\newblock {\em {Pattern Recognition}}.
\newblock Elsevier, San Diego, second edition, 2003.

\bibitem{VanderLoo2014}
M.~P.~J. Van~der Loo.
\newblock The stringdist package for approximate string matching.
\newblock {\em The R Journal}, 6(1):111--122, 2014.

\bibitem{Wang2008}
G.~Wang, Y.~Zhao, and D.~Wang.
\newblock A protein secondary structure prediction framework based on the
  extreme learning machine.
\newblock {\em Neurocomputing}, 72(1-3):262--268, 2008.

\bibitem{Wang2016}
S.~Wang, J.~Peng, J.~Ma, and J.~Xu.
\newblock Protein secondary structure prediction using deep convolutional
  neural fields.
\newblock {\em Scientific reports}, 6:18962, 2016.

\bibitem{Ward2003}
J.~J. Ward, L.~J. McGuffin, B.~F. Buxton, and D.~T. Jones.
\newblock Secondary structure prediction with support vector machines.
\newblock {\em Bioinformatics}, 19(13):1650--1655, 2003.

\bibitem{Wu2008}
X.~Wu, V.~Kumar, J.~R. Quinlan, J.~Ghosh, Q.~Yang, H.~Motoda, G.~J. McLachlan,
  A.~Ng, B.~Liu, and S.~Y. Philip.
\newblock Top 10 algorithms in data mining.
\newblock {\em Knowledge and information systems}, 14(1):1--37, 2008.

\bibitem{Yaseen2014}
A.~Yaseen and Y.~Li.
\newblock Context-based features enhance protein secondary structure prediction
  accuracy.
\newblock {\em Journal of chemical information and modeling}, 54(3):992--1002,
  2014.

\bibitem{Zamani2015}
M.~Zamani and S.~C. Kremer.
\newblock A multi-stage protein secondary structure prediction system using
  machine learning and information theory.
\newblock In {\em Bioinformatics and Biomedicine (BIBM), 2015 IEEE
  International Conference on}, pages 1304--1309. IEEE, 2015.

\bibitem{Zangooei2012}
M.~H. Zangooei and S.~Jalili.
\newblock Protein secondary structure prediction using dwkf based on
  svr-nsgaii.
\newblock {\em Neurocomputing}, 94:87--101, 2012.

\bibitem{Zangooei2012a}
M.~H. Zangooei and S.~Jalili.
\newblock Pssp with dynamic weighted kernel fusion based on svm-phgs.
\newblock {\em Knowledge-Based Systems}, 27:424--442, 2012.

\end{thebibliography}
\end{document}